\documentclass[pre,twocolumn,floatfix,superscriptaddress,a4paper,nofootinbib, amsmath,amssymb]{revtex4-1}

\usepackage[section]{placeins}

\usepackage{graphicx}
\usepackage{bm}

\usepackage{xcolor}

\newcommand{\curlyL}{\mathcal{L}}

\newcommand{\SUMIGRZERO}{\sum\limits_{i>0}}

\usepackage[utf8]{inputenc}

\usepackage{subfiles}

\begin{document}

\title{Pumping and mixing in active pores}

\author{G. C. Antunes}
 \email{g.antunes@fz-juelich.de}
\affiliation{Max--Planck--Institut f\"{u}r Intelligente Systeme, Heisenbergstr.~3, 70569 Stuttgart,\,Germany}
\affiliation{IV.\,Institut f\"ur Theoretische Physik, Universit\"{a}t Stuttgart, Pfaffenwaldring 57, 70569 Stuttgart,\,Germany}
\affiliation{Helmholtz-Institut Erlangen-Nürnberg für Erneuerbare Energien (IEK--11), Forschungszentrum J\"ulich, Cauer Str.~1, 91058 Erlangen,\,Germany}

\author{P. Malgaretti}
 \email{p.malgaretti@fz-juelich.de}
 \affiliation{Max--Planck--Institut f\"{u}r Intelligente Systeme, Heisenbergstr.~3, 70569 Stuttgart,\,Germany}
\affiliation{IV.\,Institut f\"ur Theoretische Physik, Universit\"{a}t Stuttgart, Pfaffenwaldring 57, 70569 Stuttgart,\,Germany}
\affiliation{Helmholtz-Institut Erlangen-Nürnberg für Erneuerbare Energien (IEK--11), Forschungszentrum J\"ulich, Cauer Str.~1, 91058 Erlangen,\,Germany}

\author{J. Harting}
\affiliation{Helmholtz-Institut Erlangen-Nürnberg für Erneuerbare Energien (IEK--11), Forschungszentrum J\"ulich, Cauer Str.~1, 91058 Erlangen,\,Germany}
\affiliation{Department Chemie- und Bioingenieurwesen und Department Physik, Friedrich-Alexander-Universit\"at Erlangen-N\"urnberg, F\"{u}rther Stra{\ss}e 248, 90429 N\"{u}rnberg, Germany}

\author{S. Dietrich}
\affiliation{Max--Planck--Institut f\"{u}r Intelligente Systeme, Heisenbergstr.~3, 70569 Stuttgart,\,Germany}
\affiliation{IV.\,Institut f\"ur Theoretische Physik, Universit\"{a}t Stuttgart, Pfaffenwaldring 57, 70569 Stuttgart,\,Germany}

\date{\today}

\begin{abstract}
We show both numerically and analytically that a chemically patterned active pore can act as a micro/nano-pump for fluids, even if it is fore-aft symmetric. This is possible due to a spontaneous symmetry breaking which occurs when advection rather than diffusion is the dominant mechanism of solute transport. We further demonstrate that, for pumping and tuning the flow rate, a combination of geometrical and chemical inhomogeneities is required. For certain parameter values, the flow is unsteady, and persistent oscillations with a tunable frequency appear. Finally, we find that the flow exhibits convection rolls and hence promotes mixing in the low Reynolds number regime.
\end{abstract}

\maketitle

The manipulation of fluid flow at the micro- and nanometer scale has currently attracted the attention of a large scientific community~\cite{Squires2005,Whitesides2006,Novotny2017,Hou2017}. 
Indeed, the emerging techniques of micro- and nanofluidics have been applied successfully to the synthesis of microparticles, to the transport of biomaterials, and to the functioning of chemical reactors \cite{Shestopalov2004,Amreen2021,Bailey2021}. Such techniques have been exploited in biomedical research to study and manipulate biological tissues, and to develop both new drugs and means of delivering them \cite{Wang2018,Yang2020,Ma2021,Egrov2021,Zhao2016}. Similarly, inkjet printing, a common technique for 3D fabrication, requires fluids to be pumped through channels with a diameter of $\sim 50-80 \mu m$ \cite{Warsi2018, Guo2017,lohse_fundamental_2022}.  Moreover, lab-on-a-chip setups have been employed both in medical research~\cite{Dittrich2006,Pol2017,Francesko2019} as well as in clinical diagnosis and treatment~\cite{Hou2017,Yang2020}. In all these situations, a fluid needs to be pumped in a controlled fashion, making micropumps a basic component of many microfluidic systems \cite{Wang2018,Laser2004}. In addition, controlling chemical reactions within such microfluidic devices requires the stirring of solutions by means of micromixers~\cite{Stroock2002,Cai2017}. 
As the cross-sections of the channels are reduced, surface and finite size effects become more relevant and can be exploited for designed microfluidic applications~\cite{Luo2014,Zhao2012,Gaikwad2020,Eloul2021,Abecassis2009,Tan2019}.

From this perspective, phoretic phenomena~\cite{Anderson1989} can provide an intriguing technique to manipulate fluid flows in a micro/nano-channel or pore. In particular in diffusioosmosis, inhomogeneous densities of certain components of the solution set up local pressure imbalances in the vicinity of solid walls, hence leading to the onset of a net fluid flow~\cite{Anderson1989}.
One way of inducing such local pressure gradients at steady-state is to fabricate chemically or geometrically inhomogeneous pores \cite{Yu2020,Michelin2019,Michelin2015_2}.
A similar procedure has already been exploited for colloids and led to the realization of self-phoretic Janus particles~\cite{Howse2007,Ebbens_Review,Bechinger_RMP,Juliane_review,Esplandiu2018,PopescuReview,Malgaretti2021}. For these colloids net motion is attained because half of their surface is covered with a catalyst promoting a chemical reaction which in turn is responsible for the inhomogeneous density of reaction products along the surface of the colloid. Interestingly, even colloids homogeneously covered with catalyst can swim due to an instability triggered when the transport of solute by advection is comparable to that due to diffusion \cite{Michelin2013, deBuyl2013}. 

\begin{figure}[t]
	\centering
   \includegraphics[width=0.5\textwidth]{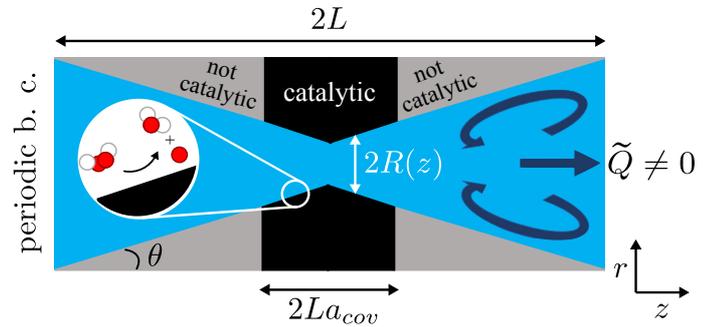}
   \caption{Longitudinal section of the axially symmetric and partially active pore with length $2L$ and variable radius $R(z)$. The decomposition of a chemical species to produce solute occurs solely in the catalytically active part of the inner pore wall with length $2La_{cov}$ (black). Convection rolls (dark blue arrows) appear due to diffusioosmosis and eventually may lead to the onset of a net nonzero flow rate $\tilde{Q}$.
   }
    	\label{fig:model}
\end{figure}

In this Letter, we show that diffusiophoresis within inhomogeneously chemically patterned pores can lead to the onset of spontaneous symmetry breaking (pumping), oscillations, and mixing. By means of both numerical simulations and analytical modeling we show that the onset of these regimes is controlled by three dimensionless parameters: the P\'eclet number (controlling the role of advection), the chemical patterning (controlling the surface inhomogeneity), and the corrugation (controlling the geometrical inhomogeneity).

In particular, as reported~\cite{Michelin2020, Chen2021} previously,  pumping does not occur for pores with homogeneous (constant) cross-sections or for pores with chemically homogeneous surface properties. In addition, beyond the stationary Stokes limit, the steady flow becomes unstable and we observe the onset of an ``inertial phoresis'' regime characterized by sustained oscillations, the frequency of which can be tuned upon varying the extent of the catalytic coverage of the pore. In all cases, convection rolls emerge, which can be exploited as so-called micromixers~\cite{Stroock2002,SNH12}.

\begin{figure*}[t]
	\centering
   \includegraphics[width=1.0\textwidth]{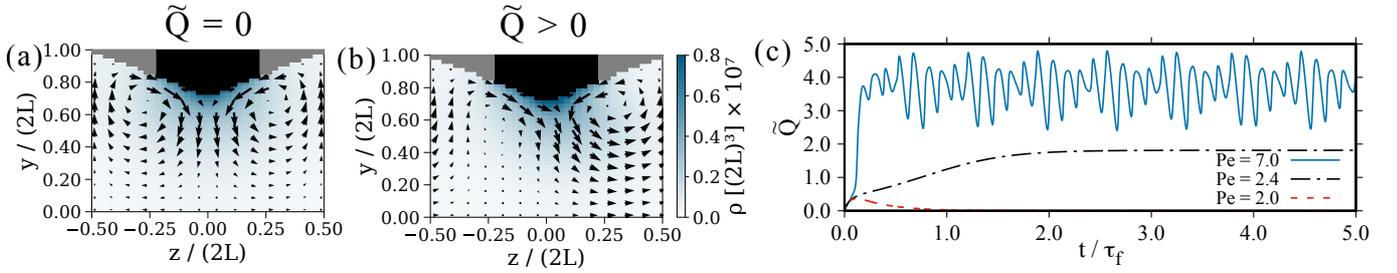}
   \caption{Panels (a) and (b). Snapshots of the steady-state velocity profile in the plane $x=0$ for $Pe=2.0$ and $2.4$, respectively. (c) Flow rate $\tilde{Q}$ as a function of time, with $a_{cov}=0.45$. The parameters are $R_{max}/(2L)=1$, $\nu \tau_f/(2L)^2= 1$, $\beta U_0 = 4 \times 10^ {-4}$, $l/(2L)=0.1$, $\xi (2L)^2\tau_f=1.5 \times 10^7$, and $\chi \tau_f = 9.6$. In lattice units: $L=20$, $\eta = 1/6$, $U_0 = 4 \times 10^ {-4}$, $l=4$, $\xi=1$, $\chi=10^{-3}$, $\beta=1$, and $\theta = \pi/6$. The simulation box is of size $80\times80\times40$. }
    	\label{fig:Q_t}
\end{figure*}
In the following, we consider hourglass-shaped pores, see Fig.~\ref{fig:model}, which is axially symmetric with respect to the $z$-axis and symmetric with respect to the plane $z=0$ in the center (fore-aft symmetry). The pore is defined by its length $2L$, its maximum radius $R_{max}$, and its opening angle $\theta$ (Fig.~\ref{fig:model}). Thus, the spatially varying radius $R(z)$ is given by
\begin{equation}
R(z) = R_{max}(\theta) - \tan(\theta)(L-|z|).
\end{equation} 
Upon a change of $\theta$, $R_{max}$ is adjusted in order to approximately conserve the volume of the pore. The pore is filled with a Newtonian fluid, the dynamics of which is governed by the Navier-Stokes equation with no-slip boundary conditions on the pore walls. The surface of the pore is patterned with a catalytic coating in the section $z \in \{- a_{cov}L, a_{cov}L\}$, where the covering fraction $a_{cov}$ can vary from zero (no coating) to one (full coating). Such a catalytic coating enables a chemical reaction resulting in the local synthesis of reaction products, which in the following are summarily called ``solute''. The solute is decomposed homogeneously in the bulk fluid with rate $\chi$ (with dimension sec$^{-1}$). In order to keep the model simple, we assume that the number densities of the reactants are kept constant in time and homogeneous in space, and that the number density $\rho$ of the solute is much smaller than the number densities of the fluid molecules such that effectively it can be regarded as an ideal gas. The effective interaction potential $U_{wall}$ between the solute molecules and the pore walls is assumed to be a piece-wise linear function of the distance $r$ from the wall,
\begin{equation}
\label{eq:potentialMain}
U_{wall}(r) = \begin{cases} 
      U_0(1 -r/l) & 0\leq r\leq l, \\
      0 & l\leq r,
   \end{cases}
\end{equation}
where $l$ is the range of $U_{wall}$. It is assumed to be much smaller than the average radius of the pore $R_0 = R_{max} - \tan(\theta)L/2$. The overdamped dynamics of the solute number density is governed by the Smoluchowski equation,
\begin{equation}
	\label{eq:rho}
	\dot{\rho} = - \nabla {\bf j} - \chi \rho,\quad 
\textbf{j}= -D\nabla\rho -\beta D\rho\nabla U_{wall}+ \bm{v}\rho,
\end{equation}
where $D$ is the diffusion coefficient of the solute, $\beta=1/(k_BT)$ is the inverse thermal energy, $\bm{v}$ is the velocity field of the solution, and $\chi$ is an empirical input parameter with the unit sec$^{-1}$. Equations~\eqref{eq:rho} obey periodic boundary conditions on the ends of the pore segment and flux boundary conditions on the surface of the pore, 
\begin{equation}\label{eq:source_sims}
\bf{j} \cdot \bf{n} |_{\text{(z, $\phi$, $r=R(z)$)}} = \begin{cases} 
      \xi, &  |z| < L a_{cov}, \\
      0, &  \text{otherwise},
   \end{cases}
\end{equation}
where $\phi$ is the azimuthal angle, $\textbf{n}$ is a unit vector perpendicular to the pore wall (pointing towards the inside of pore), and $\xi$ is a positive constant with dimension $[m^2 s]^{-1}$ .

The wall-solute interaction results in a laterally inhomogeneous pressure along the wall, hence coupling Eq.~\eqref{eq:rho} with the Navier-Stokes equation and the continuity equation for the fluid density (see Sup.~Mat.) These three equations are solved in parallel using a finite-difference solver for the first one (second-order in space and first-order in time), and the lattice Boltzmann method (LBM)~\cite{Benzi1992,Krueger_book,Harting2016} for the other two ones. (Details of the numerical implementation can be found in Ref.~\cite{Peter2019}.) \\
In the following, we report all quantities in units of the pore length $2L$ ($40$ spatial lattice units), and of the fluid relaxation time $\tau_f =(2L)^2/\nu$ ($9600$ temporal lattice units), which is the time required for momentum to diffuse across the pore in the longitudinal dimension, with $\nu$ being the kinematic viscosity ($1/6$ in lattice units).  

By following Ref.~\cite{Peter2019}, in all simulations we have used parameter values as reported in the caption of Fig. \ref{fig:Q_t}. We remark that there is a maximum value of $\theta$ for which the bottleneck of the pore is shut down. This value is obtained by numerically solving for $\theta_{max}$ such that $R(z=0;\theta=\theta_{max})=0$. For the geometry under consideration, this value amounts to $\theta_{max}~\approx~0.4\pi$. The system is initialized with the fluid at rest and with a fore-aft asymmetric density profile of the solute. Since the effective interaction potential is repulsive, the flow field resulting from the initial one is directed from the solute-poor half of the pore to the solute-rich half~\cite{Anderson1989} (see Figs.~\ref{fig:Q_t}(a) and~\ref{fig:Q_t}(b)). The competition between advective and diffusive transport is key to the dynamics we report. This competition is quantified by the P\'eclet number $Pe=v^* L /D$ which is proportional to the characteristic velocity $v^*$ and sets the ratio of the timescales of diffusive and advective transport. Since it is possible to vary $Pe$ by varying any of the three quantities, due to numerical efficiency, we varied $Pe$ by tuning the diffusion coefficient. Only the solute inside the thin region around the pore walls where $U_{wall} \neq 0$ contributes to diffusioosmosis. Therefore, we focus on the transport in this region. The characteristic velocity $v^*$ is estimated from the numerical simulations by averaging the velocity of the fluid close to the pore walls (see Sup.~Mat.) for the cases in which pumping occurs, which yields $v^*=1.3$ ($0.0056$ in lattice units).  \\
For sufficiently small values of $Pe$ the flow field relaxes to a steady state characterized by convection rolls (Fig.~\ref{fig:Q_t}(a)), which act to mix the fluid. These states are characterized also by a vanishing fluid flow rate
\begin{align}
   \tilde{Q} = \frac{\tau_f}{(2L)^3} \int_{0}^{R(z)} dr \ r  \int_0^{2\pi}d \phi \ v_z(r,\phi,z) 
\end{align}
(see the dashed line in Fig.~\ref{fig:Q_t}(c)), which we report normalized by $(2L)^3/\tau_f$. However, upon increasing the value of $Pe$, we observe a non-vanishing steady-state fluid flow rate, $\tilde{Q}\neq0$ (Figs.~\ref{fig:Q_t}(b) and~\ref{fig:Q_t}(c)). In these steady states, the advection of solute compensates for the diffusion which attempts to equilibrate the solute density in both the fore and the aft half of the pore. As the convection rolls are present also in these pumping states, the channel mixes and pumps the fluid at the same time.
\begin{figure}[t]
   \centering
   \includegraphics[width=0.5\textwidth]{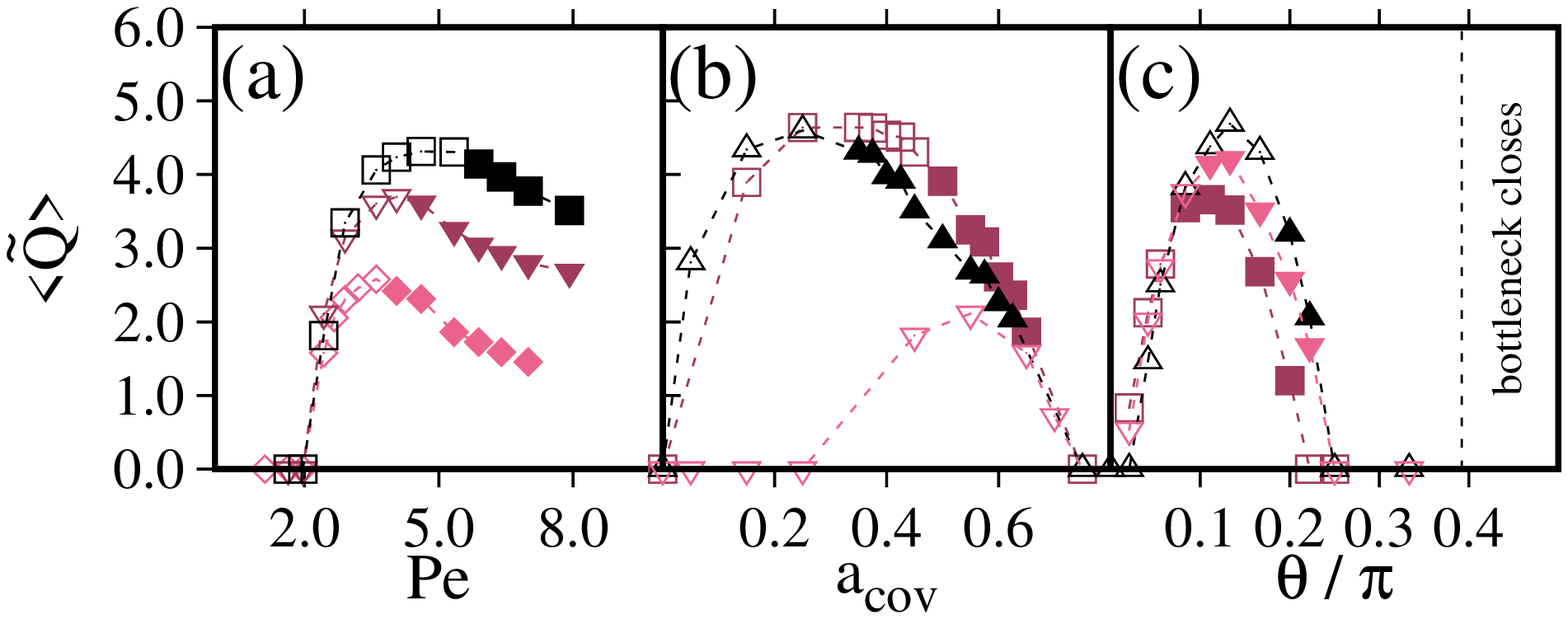}
   \caption{Time-averaged flow rate $\langle \tilde{Q} \rangle$. Open (solid) symbols mark systems which converge to a steady state (limit cycle). $\langle \tilde{Q} \rangle$ (a) as function of $Pe$ for $\theta = \pi/6$ and for $a_{cov}=\{0.45 \ (\square), \textcolor[HTML]{9E3A5E}{0.55} \ (\textcolor[HTML]{9E3A5E}{\triangledown}), \textcolor[HTML]{EA638C}{0.65} \ (\textcolor[HTML]{EA638C}{\lozenge})\}$; (b) as function of $a_{cov}$ for $\theta = \pi/6$ and for $Pe=\{\textcolor[HTML]{EA638C}{2.4} \ (\textcolor[HTML]{EA638C}{\triangledown}),\textcolor[HTML]{9E3A5E}{5.3} \ (\textcolor[HTML]{9E3A5E}{\square}),8.0\ (\triangle)\}$; (c) as function of $\theta$, for $\{Pe,a_{cov}\} = [\{5.3,0.45\}$~($\triangle$), $\{\textcolor[HTML]{EA638C}{8.0,0.45}\}$ ($\textcolor[HTML]{EA638C}{\triangledown}$), $\{\textcolor[HTML]{9E3A5E}{8.0,0.55}\}$ ($\textcolor[HTML]{9E3A5E}{\square}$)$]$. For further parameters see the caption of Fig. \ref{fig:Q_t}. In panel (c), the size of the simulation box is adjusted so as to keep the volume of the pore constant. The dashed lines are guides to the eye. 
   }
    	\label{fig:Q_D_CC}
\end{figure}

Figure~\ref{fig:Q_D_CC}(a) shows the dependence of $\tilde{Q}$ on $Pe$, and it highlights the presence of a crossover value of $Pe_c$ above which pumping (i.e., $\tilde{Q}\neq 0$) occurs. The values of both $Pe_c$ and $\tilde{Q}$ are sensitive to the chemical and geometrical properties of the pore. Indeed, Fig.~\ref{fig:Q_D_CC}(b) shows that pumping is suppressed in the limits of small ($a_{cov} \rightarrow 0$) and large (in this case $a_{cov}~\gtrsim~0.8$) chemical patterns, respectively. In particular, we have found no pumping steady state in the case of a pore fully covered with catalyst ($a_{cov}=1$). The onset of pumping is sensitive to the geometry of the pore, too. In fact, Fig.~\ref{fig:Q_D_CC}(c) shows that there are both lower and upper limits $\theta_{min}<\theta<\theta_{max}$ below and above which pumping does not occur. This is in contrast to what has been (theoretically) reported for colloidal particle which undergo a spontaneous symmetry breaking also in the case of homogeneous surface properties. 

Figure~\ref{fig:Q_D_CC} clearly shows that the three dimensionless parameters ($Pe$, $a_{cov}$, $\theta$) which we have identified play a crucial role in the onset of the spontaneous symmetry breaking. Therefore the rich phenomenology that we report here cannot be attained for low P\'eclet numbers (no advection), homogenous chemical patterning (small and large values of $a_{cov}$) and for flat or very corrugated channels (small and large values of $\theta$).

\begin{figure*}[t]
	\centering
   \includegraphics[width=1.0\textwidth]{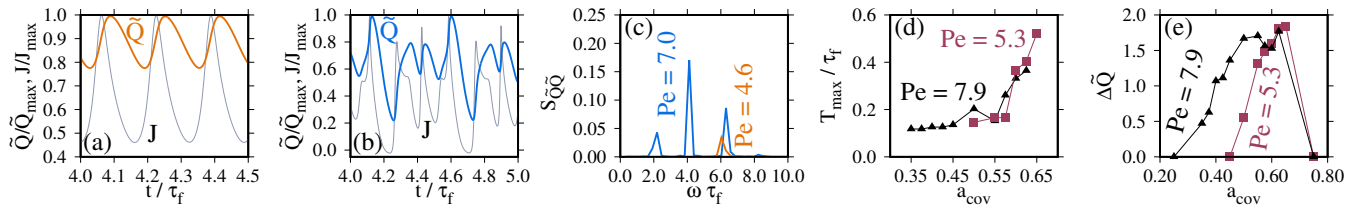}
   \caption{In panels (a) and (b) $\tilde{Q}$ and $J$ (thick lines (\textcolor[HTML]{e0730b}{orange}, \textcolor[HTML]{0b6ee0}{blue}), and \textcolor[HTML]{8e9aaf}{grey} thin lines, respectively) normalized by their maximum value. In (a) $\{a_{cov},Pe\} = \{0.55,4.6\}$ and in (b) $\{a_{cov}, Pe\} = \{0.55,7.0\}$.  (c) Power spectrum $S_{\tilde{Q}\tilde{Q}}$ of $\tilde{Q}(t)$ (from panels (a) and (b)). (d) Period of oscillations $T_{max}$ in units of $\tau_f$. (e) Amplitude of the oscillations $\Delta \tilde{Q}$. For (d) and (e), the data are shown as function of $a_{cov}$ for $Pe=\{\textcolor[HTML]{9E3A5E}{5.3}$ $(\textcolor[HTML]{9E3A5E}{\blacksquare})$, $7.9$ ($\blacktriangle$)$\}$. For further parameters see the caption of Fig. \ref{fig:Q_t}.}
    	\label{fig:freq_amp}
\end{figure*}

Remarkably, there is a regime (solid symbols in Fig.~\ref{fig:Q_D_CC}), with $Pe>Pe^{osc}$, $a_{cov}>a_{cov}^{osc}$, and $\theta > \theta^{osc}$, in which $\tilde{Q}$ exhibits sustained oscillations about a non-vanishing flow, rather than converging towards a steady state (see the full line in Fig.~\ref{fig:Q_t}(c)). These sustained oscillations are qualitatively different from those observed in Ref.~\cite{Chen2021} which occur at zero pumping rate. Here, the pulsatile-like flow arises from a negative feedback loop which works as follows. An initial increase in $\tilde{Q}$ causes the solute to be advected away from the center of the pore at a rate faster than the rate at which the catalysis at the wall can replace it. This results in a large amount of solute (which we denote as a plume) which is rapidly ejected from the wall (see Sup. Video). The depletion of solute from the center of the hourglass causes the flow $J$ of the solute, i.e.,
\begin{align}
    J=\int_{0}^{R(0)}dr \ r  \int_0^{2\pi} d \phi \ v_z(r,\phi,z=0)\rho(r,\phi,z=0) 
\end{align}
to be reduced even as $\tilde{Q}$ increases. This eventually triggers a decrease in the asymmetry of the solute between each half of the pore. Eventually, $\tilde{Q}$ is diminished, even though the center of the pore repopulates with solute, and the solute flow $J$ increases. This delay between $\tilde{Q}$ and $J$ is visible in Figs.~\ref{fig:freq_amp}(a) and (b), and is a result of the non-zero relaxation time $\tau_f=(2L)^2/\nu$ of the fluid velocity distribution. Accordingly, the sustained oscillations occur when the fluid velocity cannot adiabatically follow the solute density field. This latency triggers an instability and prevents relaxation to a steady state.

Interestingly, the onset of sustained oscillations roughly coincides with the regime in which the time-averaged flow rate $\langle \tilde{Q} \rangle$ diminishes upon an increase in $Pe$, i.e., upon favoring even further advection with respect to diffusion (Fig.~\ref{fig:Q_D_CC} (a)). Such a non-monotonic dependence of $\tilde{Q}$ on $Pe$ is reminiscent of the one observed for both isotropic and Janus colloids \cite{Michelin2013, Michelin2014, Michelin2015}. 
To further characterize the sustained oscillations of $\tilde{Q}$, we analyze their Fourier spectra ($S_{\tilde{Q}\tilde{Q}}$, see Sup.~Mat.). Figure~\ref{fig:freq_amp}(c) shows rich power spectra with multiple excited modes. In order to analyze the dependence of the power spectra on the extent $a_{cov}$ of the chemical pattern, and on $Pe$, we focus on the period $T_{max}$ associated with that frequency for which the power spectrum attains its maximum: $2 \pi \tau_f / T_{max}$. The dependence of $T_{max}$ on $a_{cov}$, normalized by the relaxation time of the fluid, is plotted in Fig.~\ref{fig:freq_amp}(d). 
In general, larger values of $a_{cov}$ result in larger values of $T_{max}$, but all values remain comparable to (but less than) the relaxation time of the fluid. Accordingly, the period of the oscillations is shorter than the relaxation time of the fluid, hence preventing the relaxation of the fluid velocity towards a steady state. Concerning the amplitude of the oscillations, Fig.~\ref{fig:freq_amp}(e) shows that larger values of $a_{cov}$ lead to larger amplitudes $\Delta \tilde{Q}= (\tilde{Q}_{max}-\tilde{Q}_{min})/2$, defined as one half of the difference between the maximum, $\tilde{Q}_{max}$, and the minimum, $\tilde{Q}_{min}$, value of $\tilde{Q}$ (i.e., to larger plumes). Interestingly, the comparison of Fig.~\ref{fig:freq_amp}(d) and Fig.~\ref{fig:freq_amp}(e) tells that larger periods $T_{max}$ are associated with larger amplitudes of the oscillations as both increase upon increasing $a_{cov}$. This different behavior on both sides of the oscillatory regime implies that the transition to oscillations from the side of smaller values of $a_{cov}$ is of a different kind as compared to the one which occurs upon approaching it from the side of larger values of $a_{cov}$. Indeed, in the former case, the amplitude of the oscillations grows smoothly from $\Delta \tilde{Q} = 0$, i.e., a supercritical Hopf bifurcation occurs~\cite{Strogatz2015}. In contrast, upon approaching from large values of $a_{cov}$ ($a_{cov}~\gtrsim~0.65$), the amplitude of the oscillations suddenly jumps from $\Delta \tilde{Q}=0$ to $\Delta \tilde{Q} \neq 0$ (i.e., oscillations in $\tilde{Q}$), i.e., a subcritical Hopf bifurcation occurs~\cite{Strogatz2015}.

\begin{figure}[t]
	\centering
   \includegraphics[width=0.5\textwidth]{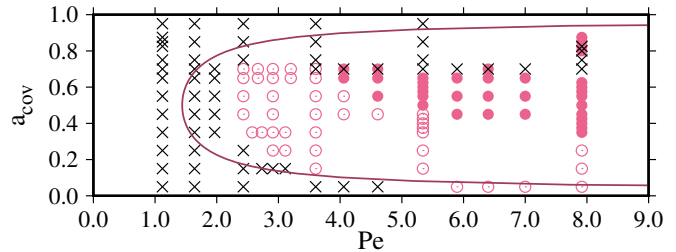}
   \caption{Classification of the asymptotic dynamics into non-pumping states ($\times$), steady pumping states ($\textcolor[HTML]{EA638C}{\circ}$), and oscillating states ($\textcolor[HTML]{EA638C}{\bullet}$). Overlapping symbols indicate bistability (see Sup.~Mat). The purple line is a semi-analytic prediction for the onset of pumping. (Concerning the parameter set, see the caption of Fig. \ref{fig:Q_t}.)}
    	\label{fig:diagram}
\end{figure}
Finally, in Fig.~\ref{fig:diagram}, we report on the asymptotic dynamics 
as a function of two out of the three dimensioless parameters identified in Fig.~\ref{fig:Q_D_CC}, namely $Pe$ and $a_{cov}$ for a given value of the corrugation $\theta=\pi/6$. 
In particular, we observe a minimum value of $Pe$ below which there is no pumping ($\tilde{Q}=0$) for any value of $a_{cov}$ and that $a_{cov} \approx 0.5$ maximizes the range of $Pe$ values for which pumping occurs. 
Interestingly, Fig.~\ref{fig:diagram} shows that not only pumping (see Fig.~\ref{fig:Q_D_CC}), but also oscillations occur for a specific range of values of $Pe$ and $a_{cov}$ (in the present case $\theta$ is fixed).
In order to understand the onset of pumping, we develop a semi-analytical approach based on the Fick-Jacobs equation~\cite{Zwanzig1992,Reguera2001,Malgaretti2013}. Within this approach, we couple Eq.~\eqref{eq:rho}, which governs the dynamics of the solute, to the stationary Stokes equation in the case of weakly varying pores for which we can apply the lubrication approximation (see Sup.~Mat.). Without any fitting parameter, the theory semi-quantitatively reproduces the onset of pumping  described by the condition $\Delta \Omega (Pe,a_{cov})=0$, where the function $\Delta \Omega(Pe, a_{cov})$ is given by the right-hand-side of Eq.~(S71) of the Sup.~Mat. . This approach captures the corresponding values of $Pe_{on}(a_{cov})$, as well as the two values of $a_{cov}^{on}$ associated with a given value of $Pe$. 

The typical experimental realization of phoresis relies on hydrogen peroxide decomposed by platinum. In such a setup, the role of the solute is played by oxygen, which has a diffusion coefficient of $D \approx 10^3 \mu m^2s^{-1}$. This setup generates flows with characteristic velocities $v^* \lesssim 10 \mu ms^{-1}$~ \cite{Ebbens2012}. According to our results, for a symmetric active pore with $a_{cov} \ \approx \ 0.5$, pumping occurs for $Pe \ \approx \ 1$, and therefore for $L \ \approx \ 10^2 - 10^3 \mu m$. The fluid relaxation time for an aqueous solution ($\nu \ \approx \ 10^6 \mu m^2 s^{-1}$) in this pore is $\tau_f \  \approx  \  10^{-2}-1 s$, and so we expect the oscillations to have a frequency in the order of $1/\tau_f \  \approx  \ 1-100$ Hz.

By means of both numerical simulations and analytical modeling, we have shown that diffusioosmosis inside pores can lead to spontaneous symmetry breaking and sustained oscillations of the flow rate. In particular, our results show that the spontaneous symmetry breaking occurs when three conditions are met simultaneously: large P\'eclet number ($Pe$ $\gtrsim$ $1$), inhomogeneous chemical patterning ($a_{cov}\neq 0,1$), and mild channel corrugation ($0<\theta<\theta_{max}$).
The oscillations, which resemble a pulsatile flow, appear as an additional instability, occurring at higher values of $Pe$ than the spontaneous symmetry breaking leading to steady pumping. They occur if the magnitude of $a_{cov}$ lies between two ``critical'' values, one showing a subcritical and the other one a supercritical Hopf bifurcation. In particular, the frequency of these oscillations can be tuned hence paving the way for the design of a phoretic microfluidic oscillator~\cite{Kim2013}. Interestingly, the  three functionalities of the active pore (mixer, pump, oscillator) can be enabled via $a_{cov}$ which may be varied by changing the light source shining on the pore in the case in which the pore is coated by a photo-activated catalyst~\cite{Palacci2014}.

\vfill

\section*{Acknowledgments}

P.M. and J.H acknowledge funding by the Deutsche Forschungsgemeinschaft (DFG, German Research Foundation) – Project-ID 416229255 – SFB 1411 and Project-ID 431791331 - SFB 1452.

\bibliography{refs}

\begin{thebibliography}{57}%
\makeatletter
\providecommand \@ifxundefined [1]{%
 \@ifx{#1\undefined}
}%
\providecommand \@ifnum [1]{%
 \ifnum #1\expandafter \@firstoftwo
 \else \expandafter \@secondoftwo
 \fi
}%
\providecommand \@ifx [1]{%
 \ifx #1\expandafter \@firstoftwo
 \else \expandafter \@secondoftwo
 \fi
}%
\providecommand \natexlab [1]{#1}%
\providecommand \enquote  [1]{``#1''}%
\providecommand \bibnamefont  [1]{#1}%
\providecommand \bibfnamefont [1]{#1}%
\providecommand \citenamefont [1]{#1}%
\providecommand \href@noop [0]{\@secondoftwo}%
\providecommand \href [0]{\begingroup \@sanitize@url \@href}%
\providecommand \@href[1]{\@@startlink{#1}\@@href}%
\providecommand \@@href[1]{\endgroup#1\@@endlink}%
\providecommand \@sanitize@url [0]{\catcode `\\12\catcode `\$12\catcode
  `\&12\catcode `\#12\catcode `\^12\catcode `\_12\catcode `\%12\relax}%
\providecommand \@@startlink[1]{}%
\providecommand \@@endlink[0]{}%
\providecommand \url  [0]{\begingroup\@sanitize@url \@url }%
\providecommand \@url [1]{\endgroup\@href {#1}{\urlprefix }}%
\providecommand \urlprefix  [0]{URL }%
\providecommand \Eprint [0]{\href }%
\providecommand \doibase [0]{http://dx.doi.org/}%
\providecommand \selectlanguage [0]{\@gobble}%
\providecommand \bibinfo  [0]{\@secondoftwo}%
\providecommand \bibfield  [0]{\@secondoftwo}%
\providecommand \translation [1]{[#1]}%
\providecommand \BibitemOpen [0]{}%
\providecommand \bibitemStop [0]{}%
\providecommand \bibitemNoStop [0]{.\EOS\space}%
\providecommand \EOS [0]{\spacefactor3000\relax}%
\providecommand \BibitemShut  [1]{\csname bibitem#1\endcsname}%
\let\auto@bib@innerbib\@empty
\bibitem [{\citenamefont {Squires}\ and\ \citenamefont
  {Quake}(2005)}]{Squires2005}%
  \BibitemOpen
  \bibfield  {author} {\bibinfo {author} {\bibfnamefont {T.~M.}\ \bibnamefont
  {Squires}}\ and\ \bibinfo {author} {\bibfnamefont {S.~R.}\ \bibnamefont
  {Quake}},\ }\href@noop {} {\bibfield  {journal} {\bibinfo  {journal} {Rev.
  Mod. Phys.}\ }\textbf {\bibinfo {volume} {77}},\ \bibinfo {pages} {977}
  (\bibinfo {year} {2005})}\BibitemShut {NoStop}%
\bibitem [{\citenamefont {Whitesides}(2006)}]{Whitesides2006}%
  \BibitemOpen
  \bibfield  {author} {\bibinfo {author} {\bibfnamefont {G.~M.}\ \bibnamefont
  {Whitesides}},\ }\href@noop {} {\bibfield  {journal} {\bibinfo  {journal}
  {Nature}\ }\textbf {\bibinfo {volume} {442}},\ \bibinfo {pages} {368}
  (\bibinfo {year} {2006})}\BibitemShut {NoStop}%
\bibitem [{\citenamefont {Novotný}\ and\ \citenamefont
  {Foret}(2017)}]{Novotny2017}%
  \BibitemOpen
  \bibfield  {author} {\bibinfo {author} {\bibfnamefont {J.}~\bibnamefont
  {Novotný}}\ and\ \bibinfo {author} {\bibfnamefont {F.}~\bibnamefont
  {Foret}},\ }\href@noop {} {\bibfield  {journal} {\bibinfo  {journal} {J. Sep.
  Sci.}\ }\textbf {\bibinfo {volume} {40}},\ \bibinfo {pages} {383} (\bibinfo
  {year} {2017})}\BibitemShut {NoStop}%
\bibitem [{\citenamefont {Hou}\ \emph {et~al.}(2017)\citenamefont {Hou},
  \citenamefont {Zhang}, \citenamefont {Santiago}, \citenamefont {Alvarez},
  \citenamefont {Ribas}, \citenamefont {Jonas}, \citenamefont {Weiss},
  \citenamefont {Andrews}, \citenamefont {Aizenberg},\ and\ \citenamefont
  {Khademhosseini}}]{Hou2017}%
  \BibitemOpen
  \bibfield  {author} {\bibinfo {author} {\bibfnamefont {X.}~\bibnamefont
  {Hou}}, \bibinfo {author} {\bibfnamefont {Y.~S.}\ \bibnamefont {Zhang}},
  \bibinfo {author} {\bibfnamefont {G.~T.}\ \bibnamefont {Santiago}}, \bibinfo
  {author} {\bibfnamefont {M.~M.}\ \bibnamefont {Alvarez}}, \bibinfo {author}
  {\bibfnamefont {J.}~\bibnamefont {Ribas}}, \bibinfo {author} {\bibfnamefont
  {S.~J.}\ \bibnamefont {Jonas}}, \bibinfo {author} {\bibfnamefont {P.~S.}\
  \bibnamefont {Weiss}}, \bibinfo {author} {\bibfnamefont {A.~M.}\ \bibnamefont
  {Andrews}}, \bibinfo {author} {\bibfnamefont {J.}~\bibnamefont {Aizenberg}},
  \ and\ \bibinfo {author} {\bibfnamefont {A.}~\bibnamefont {Khademhosseini}},\
  }\href@noop {} {\bibfield  {journal} {\bibinfo  {journal} {Nat. Rev. Mater.}\
  }\textbf {\bibinfo {volume} {2}},\ \bibinfo {pages} {17016} (\bibinfo {year}
  {2017})}\BibitemShut {NoStop}%
\bibitem [{\citenamefont {Shestopalov}\ \emph {et~al.}(2004)\citenamefont
  {Shestopalov}, \citenamefont {Tice},\ and\ \citenamefont
  {Ismagilov}}]{Shestopalov2004}%
  \BibitemOpen
  \bibfield  {author} {\bibinfo {author} {\bibfnamefont {I.}~\bibnamefont
  {Shestopalov}}, \bibinfo {author} {\bibfnamefont {J.~D.}\ \bibnamefont
  {Tice}}, \ and\ \bibinfo {author} {\bibfnamefont {R.~F.}\ \bibnamefont
  {Ismagilov}},\ }\href@noop {} {\bibfield  {journal} {\bibinfo  {journal} {Lab
  Chip}\ }\textbf {\bibinfo {volume} {4}},\ \bibinfo {pages} {316} (\bibinfo
  {year} {2004})}\BibitemShut {NoStop}%
\bibitem [{\citenamefont {Amreen}\ and\ \citenamefont
  {Goel}(2021)}]{Amreen2021}%
  \BibitemOpen
  \bibfield  {author} {\bibinfo {author} {\bibfnamefont {K.}~\bibnamefont
  {Amreen}}\ and\ \bibinfo {author} {\bibfnamefont {S.}~\bibnamefont {Goel}},\
  }\href {\doibase 10.1149/2162-8777/abdb19} {\bibfield  {journal} {\bibinfo
  {journal} {ECS J. Solid State Sci. Technol.}\ }\textbf {\bibinfo {volume}
  {10}},\ \bibinfo {pages} {017002} (\bibinfo {year} {2021})}\BibitemShut
  {NoStop}%
\bibitem [{\citenamefont {Bailey}\ \emph {et~al.}(2021)\citenamefont {Bailey},
  \citenamefont {Pinto}, \citenamefont {Hondow},\ and\ \citenamefont
  {Wu}}]{Bailey2021}%
  \BibitemOpen
  \bibfield  {author} {\bibinfo {author} {\bibfnamefont {T.}~\bibnamefont
  {Bailey}}, \bibinfo {author} {\bibfnamefont {M.}~\bibnamefont {Pinto}},
  \bibinfo {author} {\bibfnamefont {N.}~\bibnamefont {Hondow}}, \ and\ \bibinfo
  {author} {\bibfnamefont {K.-J.}\ \bibnamefont {Wu}},\ }\href@noop {}
  {\bibfield  {journal} {\bibinfo  {journal} {MethodsX}\ }\textbf {\bibinfo
  {volume} {8}},\ \bibinfo {pages} {101246} (\bibinfo {year}
  {2021})}\BibitemShut {NoStop}%
\bibitem [{\citenamefont {Wang}\ and\ \citenamefont {Fu}(2018)}]{Wang2018}%
  \BibitemOpen
  \bibfield  {author} {\bibinfo {author} {\bibfnamefont {Y.-N.}\ \bibnamefont
  {Wang}}\ and\ \bibinfo {author} {\bibfnamefont {L.-M.}\ \bibnamefont {Fu}},\
  }\href@noop {} {\bibfield  {journal} {\bibinfo  {journal} {Microelectron.
  Eng.}\ }\textbf {\bibinfo {volume} {195}},\ \bibinfo {pages} {121} (\bibinfo
  {year} {2018})}\BibitemShut {NoStop}%
\bibitem [{\citenamefont {Yang}\ \emph {et~al.}(2020)\citenamefont {Yang},
  \citenamefont {Chen}, \citenamefont {Tang}, \citenamefont {Zong},\ and\
  \citenamefont {Jiang}}]{Yang2020}%
  \BibitemOpen
  \bibfield  {author} {\bibinfo {author} {\bibfnamefont {Y.}~\bibnamefont
  {Yang}}, \bibinfo {author} {\bibfnamefont {Y.}~\bibnamefont {Chen}}, \bibinfo
  {author} {\bibfnamefont {H.}~\bibnamefont {Tang}}, \bibinfo {author}
  {\bibfnamefont {N.}~\bibnamefont {Zong}}, \ and\ \bibinfo {author}
  {\bibfnamefont {X.}~\bibnamefont {Jiang}},\ }\href@noop {} {\bibfield
  {journal} {\bibinfo  {journal} {Small Methods}\ }\textbf {\bibinfo {volume}
  {4}},\ \bibinfo {pages} {1900451} (\bibinfo {year} {2020})}\BibitemShut
  {NoStop}%
\bibitem [{\citenamefont {Ma}\ \emph {et~al.}(2021)\citenamefont {Ma},
  \citenamefont {Ma}, \citenamefont {Xu}, \citenamefont {Wang},\ and\
  \citenamefont {Sun}}]{Ma2021}%
  \BibitemOpen
  \bibfield  {author} {\bibinfo {author} {\bibfnamefont {Q.}~\bibnamefont
  {Ma}}, \bibinfo {author} {\bibfnamefont {H.}~\bibnamefont {Ma}}, \bibinfo
  {author} {\bibfnamefont {F.}~\bibnamefont {Xu}}, \bibinfo {author}
  {\bibfnamefont {X.}~\bibnamefont {Wang}}, \ and\ \bibinfo {author}
  {\bibfnamefont {W.}~\bibnamefont {Sun}},\ }\href@noop {} {\bibfield
  {journal} {\bibinfo  {journal} {Microsyst. Nanoeng.}\ }\textbf {\bibinfo
  {volume} {7}},\ \bibinfo {pages} {19} (\bibinfo {year} {2021})}\BibitemShut
  {NoStop}%
\bibitem [{\citenamefont {Egrov}\ \emph {et~al.}(2021)\citenamefont {Egrov},
  \citenamefont {Pieters}, \citenamefont {Horach-Rechtman}, \citenamefont
  {Shklover},\ and\ \citenamefont {Schroeder}}]{Egrov2021}%
  \BibitemOpen
  \bibfield  {author} {\bibinfo {author} {\bibfnamefont {E.}~\bibnamefont
  {Egrov}}, \bibinfo {author} {\bibfnamefont {C.}~\bibnamefont {Pieters}},
  \bibinfo {author} {\bibfnamefont {H.}~\bibnamefont {Horach-Rechtman}},
  \bibinfo {author} {\bibfnamefont {J.}~\bibnamefont {Shklover}}, \ and\
  \bibinfo {author} {\bibfnamefont {A.}~\bibnamefont {Schroeder}},\ }\href@noop
  {} {\bibfield  {journal} {\bibinfo  {journal} {Drug Deliv. and Transl. Res.}\
  }\textbf {\bibinfo {volume} {11}},\ \bibinfo {pages} {345} (\bibinfo {year}
  {2021})}\BibitemShut {NoStop}%
\bibitem [{\citenamefont {Zhao}\ \emph {et~al.}(2016)\citenamefont {Zhao},
  \citenamefont {Liu}, \citenamefont {Yildirimer}, \citenamefont {Zhao},
  \citenamefont {Ding}, \citenamefont {Wang}, \citenamefont {Cui},\ and\
  \citenamefont {Weitz}}]{Zhao2016}%
  \BibitemOpen
  \bibfield  {author} {\bibinfo {author} {\bibfnamefont {X.}~\bibnamefont
  {Zhao}}, \bibinfo {author} {\bibfnamefont {S.}~\bibnamefont {Liu}}, \bibinfo
  {author} {\bibfnamefont {L.}~\bibnamefont {Yildirimer}}, \bibinfo {author}
  {\bibfnamefont {H.}~\bibnamefont {Zhao}}, \bibinfo {author} {\bibfnamefont
  {R.}~\bibnamefont {Ding}}, \bibinfo {author} {\bibfnamefont {H.}~\bibnamefont
  {Wang}}, \bibinfo {author} {\bibfnamefont {W.}~\bibnamefont {Cui}}, \ and\
  \bibinfo {author} {\bibfnamefont {D.}~\bibnamefont {Weitz}},\ }\href@noop {}
  {\bibfield  {journal} {\bibinfo  {journal} {Adv. Funct. Mater.}\ }\textbf
  {\bibinfo {volume} {26}},\ \bibinfo {pages} {2809} (\bibinfo {year}
  {2016})}\BibitemShut {NoStop}%
\bibitem [{\citenamefont {Warsi}\ \emph {et~al.}(2018)\citenamefont {Warsi},
  \citenamefont {Yusuf}, \citenamefont {Al~Robaian}, \citenamefont {Khan},
  \citenamefont {Muheem},\ and\ \citenamefont {Khan}}]{Warsi2018}%
  \BibitemOpen
  \bibfield  {author} {\bibinfo {author} {\bibfnamefont {M.~H.}\ \bibnamefont
  {Warsi}}, \bibinfo {author} {\bibfnamefont {M.}~\bibnamefont {Yusuf}},
  \bibinfo {author} {\bibfnamefont {M.}~\bibnamefont {Al~Robaian}}, \bibinfo
  {author} {\bibfnamefont {M.}~\bibnamefont {Khan}}, \bibinfo {author}
  {\bibfnamefont {A.}~\bibnamefont {Muheem}}, \ and\ \bibinfo {author}
  {\bibfnamefont {S.}~\bibnamefont {Khan}},\ }\href@noop {} {\bibfield
  {journal} {\bibinfo  {journal} {Curr. Pharm. Des.}\ }\textbf {\bibinfo
  {volume} {24}},\ \bibinfo {pages} {4949} (\bibinfo {year}
  {2018})}\BibitemShut {NoStop}%
\bibitem [{\citenamefont {Guo}\ \emph {et~al.}(2017)\citenamefont {Guo},
  \citenamefont {Patanwala}, \citenamefont {Bognet},\ and\ \citenamefont
  {Ma}}]{Guo2017}%
  \BibitemOpen
  \bibfield  {author} {\bibinfo {author} {\bibfnamefont {Y.}~\bibnamefont
  {Guo}}, \bibinfo {author} {\bibfnamefont {H.}~\bibnamefont {Patanwala}},
  \bibinfo {author} {\bibfnamefont {B.}~\bibnamefont {Bognet}}, \ and\ \bibinfo
  {author} {\bibfnamefont {A.}~\bibnamefont {Ma}},\ }\href@noop {} {\bibfield
  {journal} {\bibinfo  {journal} {Rapid Prototyp. J.}\ }\textbf {\bibinfo
  {volume} {23}},\ \bibinfo {pages} {562} (\bibinfo {year} {2017})}\BibitemShut
  {NoStop}%
\bibitem [{\citenamefont {Lohse}(2022)}]{lohse_fundamental_2022}%
  \BibitemOpen
  \bibfield  {author} {\bibinfo {author} {\bibfnamefont {D.}~\bibnamefont
  {Lohse}},\ }\href@noop {} {\bibfield  {journal} {\bibinfo  {journal} {Ann.
  Rev. Fluid Mech.}\ }\textbf {\bibinfo {volume} {54}},\ \bibinfo {pages} {349}
  (\bibinfo {year} {2022})}\BibitemShut {NoStop}%
\bibitem [{\citenamefont {Dittrich}\ and\ \citenamefont
  {Manz}(2006)}]{Dittrich2006}%
  \BibitemOpen
  \bibfield  {author} {\bibinfo {author} {\bibfnamefont {P.~S.}\ \bibnamefont
  {Dittrich}}\ and\ \bibinfo {author} {\bibfnamefont {A.}~\bibnamefont
  {Manz}},\ }\href@noop {} {\bibfield  {journal} {\bibinfo  {journal} {Nat.
  Rev. Drug Discov.}\ }\textbf {\bibinfo {volume} {5}},\ \bibinfo {pages} {210}
  (\bibinfo {year} {2006})}\BibitemShut {NoStop}%
\bibitem [{\citenamefont {Pol}\ \emph {et~al.}(2017)\citenamefont {Pol},
  \citenamefont {Céspedes}, \citenamefont {Gabriel},\ and\ \citenamefont
  {Baeza}}]{Pol2017}%
  \BibitemOpen
  \bibfield  {author} {\bibinfo {author} {\bibfnamefont {R.}~\bibnamefont
  {Pol}}, \bibinfo {author} {\bibfnamefont {F.}~\bibnamefont {Céspedes}},
  \bibinfo {author} {\bibfnamefont {D.}~\bibnamefont {Gabriel}}, \ and\
  \bibinfo {author} {\bibfnamefont {M.}~\bibnamefont {Baeza}},\ }\href@noop {}
  {\bibfield  {journal} {\bibinfo  {journal} {Trends Anal. Chem.}\ }\textbf
  {\bibinfo {volume} {95}},\ \bibinfo {pages} {62 } (\bibinfo {year}
  {2017})}\BibitemShut {NoStop}%
\bibitem [{\citenamefont {Francesko}\ \emph {et~al.}(2019)\citenamefont
  {Francesko}, \citenamefont {Cardoso},\ and\ \citenamefont
  {Lanceros-Méndez}}]{Francesko2019}%
  \BibitemOpen
  \bibfield  {author} {\bibinfo {author} {\bibfnamefont {A.}~\bibnamefont
  {Francesko}}, \bibinfo {author} {\bibfnamefont {V.~F.}\ \bibnamefont
  {Cardoso}}, \ and\ \bibinfo {author} {\bibfnamefont {S.}~\bibnamefont
  {Lanceros-Méndez}},\ }\href@noop {} {\emph {\bibinfo {title}
  {\normalfont{in} \textit{Microfluidics for Pharmaceutical Applications: From
  Nano/Micro Systems Fabrication to Controlled Drug Delivery}}}},\ edited by\
  \bibinfo {editor} {\bibfnamefont {H.~A.}\ \bibnamefont {Santos}}, \bibinfo
  {editor} {\bibfnamefont {D.}~\bibnamefont {Liu}}, \ and\ \bibinfo {editor}
  {\bibfnamefont {H.}~\bibnamefont {Zhang}}\ (\bibinfo  {publisher} {William
  Andrew Publishing},\ \bibinfo {address} {Norwich},\ \bibinfo {year} {2019})\
  p.~\bibinfo {pages} {3}\BibitemShut {NoStop}%
\bibitem [{\citenamefont {Laser}\ and\ \citenamefont
  {Santiago}(2004)}]{Laser2004}%
  \BibitemOpen
  \bibfield  {author} {\bibinfo {author} {\bibfnamefont {D.~J.}\ \bibnamefont
  {Laser}}\ and\ \bibinfo {author} {\bibfnamefont {J.~G.}\ \bibnamefont
  {Santiago}},\ }\href@noop {} {\bibfield  {journal} {\bibinfo  {journal} {J.
  Micromech. Microeng.}\ }\textbf {\bibinfo {volume} {14}},\ \bibinfo {pages}
  {R35} (\bibinfo {year} {2004})}\BibitemShut {NoStop}%
\bibitem [{\citenamefont {Stroock}\ \emph {et~al.}(2002)\citenamefont
  {Stroock}, \citenamefont {Dertinger}, \citenamefont {Ajdari}, \citenamefont
  {Mezi{\'c}}, \citenamefont {Stone},\ and\ \citenamefont
  {Whitesides}}]{Stroock2002}%
  \BibitemOpen
  \bibfield  {author} {\bibinfo {author} {\bibfnamefont {A.~D.}\ \bibnamefont
  {Stroock}}, \bibinfo {author} {\bibfnamefont {S.~K.~W.}\ \bibnamefont
  {Dertinger}}, \bibinfo {author} {\bibfnamefont {A.}~\bibnamefont {Ajdari}},
  \bibinfo {author} {\bibfnamefont {I.}~\bibnamefont {Mezi{\'c}}}, \bibinfo
  {author} {\bibfnamefont {H.~A.}\ \bibnamefont {Stone}}, \ and\ \bibinfo
  {author} {\bibfnamefont {G.~M.}\ \bibnamefont {Whitesides}},\ }\href@noop {}
  {\bibfield  {journal} {\bibinfo  {journal} {Science}\ }\textbf {\bibinfo
  {volume} {295}},\ \bibinfo {pages} {647} (\bibinfo {year}
  {2002})}\BibitemShut {NoStop}%
\bibitem [{\citenamefont {Cai}\ \emph {et~al.}(2017)\citenamefont {Cai},
  \citenamefont {Xue}, \citenamefont {Zhang},\ and\ \citenamefont
  {Lin}}]{Cai2017}%
  \BibitemOpen
  \bibfield  {author} {\bibinfo {author} {\bibfnamefont {G.}~\bibnamefont
  {Cai}}, \bibinfo {author} {\bibfnamefont {L.}~\bibnamefont {Xue}}, \bibinfo
  {author} {\bibfnamefont {H.}~\bibnamefont {Zhang}}, \ and\ \bibinfo {author}
  {\bibfnamefont {J.}~\bibnamefont {Lin}},\ }\href@noop {} {\bibfield
  {journal} {\bibinfo  {journal} {Micromachines}\ }\textbf {\bibinfo {volume}
  {8}},\ \bibinfo {pages} {9} (\bibinfo {year} {2017})}\BibitemShut {NoStop}%
\bibitem [{\citenamefont {Luo}\ \emph {et~al.}(2014)\citenamefont {Luo},
  \citenamefont {Holden},\ and\ \citenamefont {White}}]{Luo2014}%
  \BibitemOpen
  \bibfield  {author} {\bibinfo {author} {\bibfnamefont {L.}~\bibnamefont
  {Luo}}, \bibinfo {author} {\bibfnamefont {D.~A.}\ \bibnamefont {Holden}}, \
  and\ \bibinfo {author} {\bibfnamefont {H.~S.}\ \bibnamefont {White}},\
  }\href@noop {} {\bibfield  {journal} {\bibinfo  {journal} {ACS Nano}\
  }\textbf {\bibinfo {volume} {8}},\ \bibinfo {pages} {3023} (\bibinfo {year}
  {2014})}\BibitemShut {NoStop}%
\bibitem [{\citenamefont {Zhao}\ and\ \citenamefont {Yang}(2012)}]{Zhao2012}%
  \BibitemOpen
  \bibfield  {author} {\bibinfo {author} {\bibfnamefont {C.}~\bibnamefont
  {Zhao}}\ and\ \bibinfo {author} {\bibfnamefont {C.}~\bibnamefont {Yang}},\
  }\href@noop {} {\bibfield  {journal} {\bibinfo  {journal} {Microfluid.
  Nanofluid.}\ }\textbf {\bibinfo {volume} {13}},\ \bibinfo {pages} {179}
  (\bibinfo {year} {2012})}\BibitemShut {NoStop}%
\bibitem [{\citenamefont {Gaikwad}\ \emph {et~al.}(2020)\citenamefont
  {Gaikwad}, \citenamefont {Kumar},\ and\ \citenamefont
  {Mondal}}]{Gaikwad2020}%
  \BibitemOpen
  \bibfield  {author} {\bibinfo {author} {\bibfnamefont {H.~S.}\ \bibnamefont
  {Gaikwad}}, \bibinfo {author} {\bibfnamefont {G.}~\bibnamefont {Kumar}}, \
  and\ \bibinfo {author} {\bibfnamefont {P.~K.}\ \bibnamefont {Mondal}},\
  }\href@noop {} {\bibfield  {journal} {\bibinfo  {journal} {Soft Matter}\
  }\textbf {\bibinfo {volume} {16}},\ \bibinfo {pages} {6304} (\bibinfo {year}
  {2020})}\BibitemShut {NoStop}%
\bibitem [{\citenamefont {Eloul}\ and\ \citenamefont
  {Frenkel}(2021)}]{Eloul2021}%
  \BibitemOpen
  \bibfield  {author} {\bibinfo {author} {\bibfnamefont {S.}~\bibnamefont
  {Eloul}}\ and\ \bibinfo {author} {\bibfnamefont {D.}~\bibnamefont
  {Frenkel}},\ }\href@noop {} {\bibfield  {journal} {\bibinfo  {journal} {Soft
  Matter}\ }\textbf {\bibinfo {volume} {17}},\ \bibinfo {pages} {1173}
  (\bibinfo {year} {2021})}\BibitemShut {NoStop}%
\bibitem [{\citenamefont {Ab{\'e}cassis}\ \emph {et~al.}(2009)\citenamefont
  {Ab{\'e}cassis}, \citenamefont {Cottin-Bizonne}, \citenamefont {Ybert},
  \citenamefont {Ajdari},\ and\ \citenamefont {Bocquet}}]{Abecassis2009}%
  \BibitemOpen
  \bibfield  {author} {\bibinfo {author} {\bibfnamefont {B.}~\bibnamefont
  {Ab{\'e}cassis}}, \bibinfo {author} {\bibfnamefont {C.}~\bibnamefont
  {Cottin-Bizonne}}, \bibinfo {author} {\bibfnamefont {C.}~\bibnamefont
  {Ybert}}, \bibinfo {author} {\bibfnamefont {A.}~\bibnamefont {Ajdari}}, \
  and\ \bibinfo {author} {\bibfnamefont {L.}~\bibnamefont {Bocquet}},\
  }\href@noop {} {\bibfield  {journal} {\bibinfo  {journal} {New J. Phys.}\
  }\textbf {\bibinfo {volume} {11}},\ \bibinfo {pages} {075022} (\bibinfo
  {year} {2009})}\BibitemShut {NoStop}%
\bibitem [{\citenamefont {Tan}\ \emph {et~al.}(2019)\citenamefont {Tan},
  \citenamefont {Yang},\ and\ \citenamefont {Ripoll}}]{Tan2019}%
  \BibitemOpen
  \bibfield  {author} {\bibinfo {author} {\bibfnamefont {Z.}~\bibnamefont
  {Tan}}, \bibinfo {author} {\bibfnamefont {M.}~\bibnamefont {Yang}}, \ and\
  \bibinfo {author} {\bibfnamefont {M.}~\bibnamefont {Ripoll}},\ }\href@noop {}
  {\bibfield  {journal} {\bibinfo  {journal} {Phys. Rev. Applied}\ }\textbf
  {\bibinfo {volume} {11}},\ \bibinfo {pages} {054004} (\bibinfo {year}
  {2019})}\BibitemShut {NoStop}%
\bibitem [{\citenamefont {Anderson}(1989)}]{Anderson1989}%
  \BibitemOpen
  \bibfield  {author} {\bibinfo {author} {\bibfnamefont {J.~L.}\ \bibnamefont
  {Anderson}},\ }\href@noop {} {\bibfield  {journal} {\bibinfo  {journal} {Ann.
  Rev. Fluid Mech.}\ }\textbf {\bibinfo {volume} {21}},\ \bibinfo {pages}
  {061701} (\bibinfo {year} {1989})}\BibitemShut {NoStop}%
\bibitem [{\citenamefont {Yu}\ \emph {et~al.}(2020)\citenamefont {Yu},
  \citenamefont {Athanassiadis}, \citenamefont {Popescu}, \citenamefont
  {Chikkadi}, \citenamefont {G\"uth}, \citenamefont {Singh}, \citenamefont
  {Qiu},\ and\ \citenamefont {Fischer}}]{Yu2020}%
  \BibitemOpen
  \bibfield  {author} {\bibinfo {author} {\bibfnamefont {T.}~\bibnamefont
  {Yu}}, \bibinfo {author} {\bibfnamefont {A.~G.}\ \bibnamefont
  {Athanassiadis}}, \bibinfo {author} {\bibfnamefont {M.~N.}\ \bibnamefont
  {Popescu}}, \bibinfo {author} {\bibfnamefont {V.}~\bibnamefont {Chikkadi}},
  \bibinfo {author} {\bibfnamefont {A.}~\bibnamefont {G\"uth}}, \bibinfo
  {author} {\bibfnamefont {D.~P.}\ \bibnamefont {Singh}}, \bibinfo {author}
  {\bibfnamefont {T.}~\bibnamefont {Qiu}}, \ and\ \bibinfo {author}
  {\bibfnamefont {P.}~\bibnamefont {Fischer}},\ }\href@noop {} {\bibfield
  {journal} {\bibinfo  {journal} {ACS Nano}\ }\textbf {\bibinfo {volume}
  {14}},\ \bibinfo {pages} {13673} (\bibinfo {year} {2020})}\BibitemShut
  {NoStop}%
\bibitem [{\citenamefont {Michelin}\ and\ \citenamefont
  {Lauga}(2019)}]{Michelin2019}%
  \BibitemOpen
  \bibfield  {author} {\bibinfo {author} {\bibfnamefont {S.}~\bibnamefont
  {Michelin}}\ and\ \bibinfo {author} {\bibfnamefont {E.}~\bibnamefont
  {Lauga}},\ }\href@noop {} {\bibfield  {journal} {\bibinfo  {journal} {Sci.
  Rep.}\ }\textbf {\bibinfo {volume} {9}},\ \bibinfo {pages} {10788} (\bibinfo
  {year} {2019})}\BibitemShut {NoStop}%
\bibitem [{\citenamefont {Michelin}\ \emph {et~al.}(2015)\citenamefont
  {Michelin}, \citenamefont {Montenegro-Johnson}, \citenamefont {De~Canio},
  \citenamefont {Lobato-Dauzier},\ and\ \citenamefont
  {Lauga}}]{Michelin2015_2}%
  \BibitemOpen
  \bibfield  {author} {\bibinfo {author} {\bibfnamefont {S.}~\bibnamefont
  {Michelin}}, \bibinfo {author} {\bibfnamefont {T.~D.}\ \bibnamefont
  {Montenegro-Johnson}}, \bibinfo {author} {\bibfnamefont {G.}~\bibnamefont
  {De~Canio}}, \bibinfo {author} {\bibfnamefont {N.}~\bibnamefont
  {Lobato-Dauzier}}, \ and\ \bibinfo {author} {\bibfnamefont {E.}~\bibnamefont
  {Lauga}},\ }\href@noop {} {\bibfield  {journal} {\bibinfo  {journal} {Soft
  Matter}\ }\textbf {\bibinfo {volume} {11}},\ \bibinfo {pages} {5804}
  (\bibinfo {year} {2015})}\BibitemShut {NoStop}%
\bibitem [{\citenamefont {Howse}\ \emph {et~al.}(2007)\citenamefont {Howse},
  \citenamefont {Jones}, \citenamefont {Ryan}, \citenamefont {Gough},
  \citenamefont {Vafabakhsh},\ and\ \citenamefont {Golestanian}}]{Howse2007}%
  \BibitemOpen
  \bibfield  {author} {\bibinfo {author} {\bibfnamefont {J.~R.}\ \bibnamefont
  {Howse}}, \bibinfo {author} {\bibfnamefont {R.~A.~L.}\ \bibnamefont {Jones}},
  \bibinfo {author} {\bibfnamefont {A.~J.}\ \bibnamefont {Ryan}}, \bibinfo
  {author} {\bibfnamefont {T.}~\bibnamefont {Gough}}, \bibinfo {author}
  {\bibfnamefont {R.}~\bibnamefont {Vafabakhsh}}, \ and\ \bibinfo {author}
  {\bibfnamefont {R.}~\bibnamefont {Golestanian}},\ }\href@noop {} {\bibfield
  {journal} {\bibinfo  {journal} {Phys. Rev. Lett.}\ }\textbf {\bibinfo
  {volume} {99}},\ \bibinfo {pages} {048102} (\bibinfo {year}
  {2007})}\BibitemShut {NoStop}%
\bibitem [{\citenamefont {Ebbens}\ and\ \citenamefont
  {Howse}(2010)}]{Ebbens_Review}%
  \BibitemOpen
  \bibfield  {author} {\bibinfo {author} {\bibfnamefont {S.~J.}\ \bibnamefont
  {Ebbens}}\ and\ \bibinfo {author} {\bibfnamefont {J.~R.}\ \bibnamefont
  {Howse}},\ }\href@noop {} {\bibfield  {journal} {\bibinfo  {journal} {Soft
  Matter}\ }\textbf {\bibinfo {volume} {6}},\ \bibinfo {pages} {726} (\bibinfo
  {year} {2010})}\BibitemShut {NoStop}%
\bibitem [{\citenamefont {Bechinger}\ \emph {et~al.}(2016)\citenamefont
  {Bechinger}, \citenamefont {Di~Leonardo}, \citenamefont {L\"owen},
  \citenamefont {Reichhardt}, \citenamefont {Volpe},\ and\ \citenamefont
  {Volpe}}]{Bechinger_RMP}%
  \BibitemOpen
  \bibfield  {author} {\bibinfo {author} {\bibfnamefont {C.}~\bibnamefont
  {Bechinger}}, \bibinfo {author} {\bibfnamefont {R.}~\bibnamefont
  {Di~Leonardo}}, \bibinfo {author} {\bibfnamefont {H.}~\bibnamefont
  {L\"owen}}, \bibinfo {author} {\bibfnamefont {C.}~\bibnamefont {Reichhardt}},
  \bibinfo {author} {\bibfnamefont {G.}~\bibnamefont {Volpe}}, \ and\ \bibinfo
  {author} {\bibfnamefont {G.}~\bibnamefont {Volpe}},\ }\href@noop {}
  {\bibfield  {journal} {\bibinfo  {journal} {Rev. Mod. Phys.}\ }\textbf
  {\bibinfo {volume} {88}},\ \bibinfo {pages} {045006} (\bibinfo {year}
  {2016})}\BibitemShut {NoStop}%
\bibitem [{\citenamefont {Safdar}\ \emph {et~al.}(2017)\citenamefont {Safdar},
  \citenamefont {Simmchen},\ and\ \citenamefont {Jänis}}]{Juliane_review}%
  \BibitemOpen
  \bibfield  {author} {\bibinfo {author} {\bibfnamefont {M.}~\bibnamefont
  {Safdar}}, \bibinfo {author} {\bibfnamefont {J.}~\bibnamefont {Simmchen}}, \
  and\ \bibinfo {author} {\bibfnamefont {J.}~\bibnamefont {Jänis}},\
  }\href@noop {} {\bibfield  {journal} {\bibinfo  {journal} {Environ. Sci.:
  Nano}\ }\textbf {\bibinfo {volume} {4}},\ \bibinfo {pages} {1602} (\bibinfo
  {year} {2017})}\BibitemShut {NoStop}%
\bibitem [{\citenamefont {Esplandiu}\ \emph {et~al.}(2018)\citenamefont
  {Esplandiu}, \citenamefont {Zhang}, \citenamefont {Fraxedas}, \citenamefont
  {Sepulveda},\ and\ \citenamefont {Reguera}}]{Esplandiu2018}%
  \BibitemOpen
  \bibfield  {author} {\bibinfo {author} {\bibfnamefont {M.~J.}\ \bibnamefont
  {Esplandiu}}, \bibinfo {author} {\bibfnamefont {K.}~\bibnamefont {Zhang}},
  \bibinfo {author} {\bibfnamefont {J.}~\bibnamefont {Fraxedas}}, \bibinfo
  {author} {\bibfnamefont {B.}~\bibnamefont {Sepulveda}}, \ and\ \bibinfo
  {author} {\bibfnamefont {D.}~\bibnamefont {Reguera}},\ }\href@noop {}
  {\bibfield  {journal} {\bibinfo  {journal} {Acc. Chem. Res.}\ }\textbf
  {\bibinfo {volume} {51}},\ \bibinfo {pages} {1921} (\bibinfo {year}
  {2018})}\BibitemShut {NoStop}%
\bibitem [{\citenamefont {Popescu}\ \emph {et~al.}(2018)\citenamefont
  {Popescu}, \citenamefont {Uspal}, \citenamefont {Domínguez},\ and\
  \citenamefont {Dietrich}}]{PopescuReview}%
  \BibitemOpen
  \bibfield  {author} {\bibinfo {author} {\bibfnamefont {M.~N.}\ \bibnamefont
  {Popescu}}, \bibinfo {author} {\bibfnamefont {W.~E.}\ \bibnamefont {Uspal}},
  \bibinfo {author} {\bibfnamefont {A.}~\bibnamefont {Domínguez}}, \ and\
  \bibinfo {author} {\bibfnamefont {S.}~\bibnamefont {Dietrich}},\ }\href@noop
  {} {\bibfield  {journal} {\bibinfo  {journal} {Acc. Chem. Res.}\ }\textbf
  {\bibinfo {volume} {51}},\ \bibinfo {pages} {2991} (\bibinfo {year}
  {2018})}\BibitemShut {NoStop}%
\bibitem [{\citenamefont {Malgaretti}\ and\ \citenamefont
  {Harting}(2021)}]{Malgaretti2021}%
  \BibitemOpen
  \bibfield  {author} {\bibinfo {author} {\bibfnamefont {P.}~\bibnamefont
  {Malgaretti}}\ and\ \bibinfo {author} {\bibfnamefont {J.}~\bibnamefont
  {Harting}},\ }\href@noop {} {\bibfield  {journal} {\bibinfo  {journal} {Chem.
  Nano. Mat., in press}\ } (\bibinfo {year} {2021})}\BibitemShut {NoStop}%
\bibitem [{\citenamefont {Michelin}\ \emph {et~al.}(2013)\citenamefont
  {Michelin}, \citenamefont {Lauga},\ and\ \citenamefont
  {Bartolo}}]{Michelin2013}%
  \BibitemOpen
  \bibfield  {author} {\bibinfo {author} {\bibfnamefont {S.}~\bibnamefont
  {Michelin}}, \bibinfo {author} {\bibfnamefont {E.}~\bibnamefont {Lauga}}, \
  and\ \bibinfo {author} {\bibfnamefont {D.}~\bibnamefont {Bartolo}},\
  }\href@noop {} {\bibfield  {journal} {\bibinfo  {journal} {Phys. Fluids}\
  }\textbf {\bibinfo {volume} {25}},\ \bibinfo {pages} {061701} (\bibinfo
  {year} {2013})}\BibitemShut {NoStop}%
\bibitem [{\citenamefont {de~Buyl}\ \emph {et~al.}(2013)\citenamefont
  {de~Buyl}, \citenamefont {Mikhailov},\ and\ \citenamefont
  {Kapral}}]{deBuyl2013}%
  \BibitemOpen
  \bibfield  {author} {\bibinfo {author} {\bibfnamefont {P.}~\bibnamefont
  {de~Buyl}}, \bibinfo {author} {\bibfnamefont {A.~S.}\ \bibnamefont
  {Mikhailov}}, \ and\ \bibinfo {author} {\bibfnamefont {R.}~\bibnamefont
  {Kapral}},\ }\href {\doibase 10.1209/0295-5075/103/60009} {\bibfield
  {journal} {\bibinfo  {journal} {EPL}\ }\textbf {\bibinfo {volume} {103}},\
  \bibinfo {pages} {60009} (\bibinfo {year} {2013})}\BibitemShut {NoStop}%
\bibitem [{\citenamefont {Michelin}\ \emph {et~al.}(2020)\citenamefont
  {Michelin}, \citenamefont {Game}, \citenamefont {Lauga}, \citenamefont
  {Keaveny},\ and\ \citenamefont {Papageorgiou}}]{Michelin2020}%
  \BibitemOpen
  \bibfield  {author} {\bibinfo {author} {\bibfnamefont {S.}~\bibnamefont
  {Michelin}}, \bibinfo {author} {\bibfnamefont {S.}~\bibnamefont {Game}},
  \bibinfo {author} {\bibfnamefont {E.}~\bibnamefont {Lauga}}, \bibinfo
  {author} {\bibfnamefont {E.}~\bibnamefont {Keaveny}}, \ and\ \bibinfo
  {author} {\bibfnamefont {D.}~\bibnamefont {Papageorgiou}},\ }\href@noop {}
  {\bibfield  {journal} {\bibinfo  {journal} {Soft Matter}\ }\textbf {\bibinfo
  {volume} {16}},\ \bibinfo {pages} {1259} (\bibinfo {year}
  {2020})}\BibitemShut {NoStop}%
\bibitem [{\citenamefont {Chen}\ \emph {et~al.}(2021)\citenamefont {Chen},
  \citenamefont {Chong}, \citenamefont {Liu}, \citenamefont {Verzicco},\ and\
  \citenamefont {Lohse}}]{Chen2021}%
  \BibitemOpen
  \bibfield  {author} {\bibinfo {author} {\bibfnamefont {Y.}~\bibnamefont
  {Chen}}, \bibinfo {author} {\bibfnamefont {K.~L.}\ \bibnamefont {Chong}},
  \bibinfo {author} {\bibfnamefont {L.}~\bibnamefont {Liu}}, \bibinfo {author}
  {\bibfnamefont {R.}~\bibnamefont {Verzicco}}, \ and\ \bibinfo {author}
  {\bibfnamefont {D.}~\bibnamefont {Lohse}},\ }\href {\doibase
  10.1017/jfm.2021.370} {\bibfield  {journal} {\bibinfo  {journal} {J. Fluid
  Mech.}\ }\textbf {\bibinfo {volume} {919}},\ \bibinfo {pages} {A10} (\bibinfo
  {year} {2021})}\BibitemShut {NoStop}%
\bibitem [{\citenamefont {Sarkar}\ \emph {et~al.}(2012)\citenamefont {Sarkar},
  \citenamefont {Narvaez~Salazar},\ and\ \citenamefont {Harting}}]{SNH12}%
  \BibitemOpen
  \bibfield  {author} {\bibinfo {author} {\bibfnamefont {A.}~\bibnamefont
  {Sarkar}}, \bibinfo {author} {\bibfnamefont {A.}~\bibnamefont
  {Narvaez~Salazar}}, \ and\ \bibinfo {author} {\bibfnamefont {J.}~\bibnamefont
  {Harting}},\ }\href@noop {} {\bibfield  {journal} {\bibinfo  {journal}
  {Microfluidics and Nanofluidics}\ }\textbf {\bibinfo {volume} {13}},\
  \bibinfo {pages} {19} (\bibinfo {year} {2012})}\BibitemShut {NoStop}%
\bibitem [{\citenamefont {Benzi}\ \emph {et~al.}(1992)\citenamefont {Benzi},
  \citenamefont {Succi},\ and\ \citenamefont {Vergassola}}]{Benzi1992}%
  \BibitemOpen
  \bibfield  {author} {\bibinfo {author} {\bibfnamefont {R.}~\bibnamefont
  {Benzi}}, \bibinfo {author} {\bibfnamefont {S.}~\bibnamefont {Succi}}, \ and\
  \bibinfo {author} {\bibfnamefont {M.}~\bibnamefont {Vergassola}},\
  }\href@noop {} {\bibfield  {journal} {\bibinfo  {journal} {Phys. Rep.}\
  }\textbf {\bibinfo {volume} {222}},\ \bibinfo {pages} {145} (\bibinfo {year}
  {1992})}\BibitemShut {NoStop}%
\bibitem [{\citenamefont {Kr\"uger}\ \emph {et~al.}(2017)\citenamefont
  {Kr\"uger}, \citenamefont {Kusumaatmaja}, \citenamefont {Kuzmin},
  \citenamefont {Shardt}, \citenamefont {Silva},\ and\ \citenamefont
  {Viggen}}]{Krueger_book}%
  \BibitemOpen
  \bibfield  {author} {\bibinfo {author} {\bibfnamefont {T.}~\bibnamefont
  {Kr\"uger}}, \bibinfo {author} {\bibfnamefont {H.}~\bibnamefont
  {Kusumaatmaja}}, \bibinfo {author} {\bibfnamefont {A.}~\bibnamefont
  {Kuzmin}}, \bibinfo {author} {\bibfnamefont {O.}~\bibnamefont {Shardt}},
  \bibinfo {author} {\bibfnamefont {G.}~\bibnamefont {Silva}}, \ and\ \bibinfo
  {author} {\bibfnamefont {E.}~\bibnamefont {Viggen}},\ }\href@noop {} {\emph
  {\bibinfo {title} {The Lattice Boltzmann Method}}}\ (\bibinfo  {publisher}
  {Springer},\ \bibinfo {address} {Berlin},\ \bibinfo {year}
  {2017})\BibitemShut {NoStop}%
\bibitem [{\citenamefont {Liu}\ \emph {et~al.}(2016)\citenamefont {Liu},
  \citenamefont {Kang}, \citenamefont {Leonardi}, \citenamefont {Schmieschek},
  \citenamefont {Narv{\'a}ez}, \citenamefont {Jones}, \citenamefont {Williams},
  \citenamefont {Valocchi},\ and\ \citenamefont {Harting}}]{Harting2016}%
  \BibitemOpen
  \bibfield  {author} {\bibinfo {author} {\bibfnamefont {H.}~\bibnamefont
  {Liu}}, \bibinfo {author} {\bibfnamefont {Q.}~\bibnamefont {Kang}}, \bibinfo
  {author} {\bibfnamefont {C.~R.}\ \bibnamefont {Leonardi}}, \bibinfo {author}
  {\bibfnamefont {S.}~\bibnamefont {Schmieschek}}, \bibinfo {author}
  {\bibfnamefont {A.}~\bibnamefont {Narv{\'a}ez}}, \bibinfo {author}
  {\bibfnamefont {B.~D.}\ \bibnamefont {Jones}}, \bibinfo {author}
  {\bibfnamefont {J.~R.}\ \bibnamefont {Williams}}, \bibinfo {author}
  {\bibfnamefont {A.~J.}\ \bibnamefont {Valocchi}}, \ and\ \bibinfo {author}
  {\bibfnamefont {J.}~\bibnamefont {Harting}},\ }\href@noop {} {\bibfield
  {journal} {\bibinfo  {journal} {Comput. Geosci.}\ }\textbf {\bibinfo {volume}
  {20}},\ \bibinfo {pages} {777} (\bibinfo {year} {2016})}\BibitemShut
  {NoStop}%
\bibitem [{\citenamefont {Peter}\ \emph {et~al.}(2020)\citenamefont {Peter},
  \citenamefont {Malgaretti}, \citenamefont {Rivas}, \citenamefont
  {Scagliarini}, \citenamefont {Harting},\ and\ \citenamefont
  {Dietrich}}]{Peter2019}%
  \BibitemOpen
  \bibfield  {author} {\bibinfo {author} {\bibfnamefont {T.}~\bibnamefont
  {Peter}}, \bibinfo {author} {\bibfnamefont {P.}~\bibnamefont {Malgaretti}},
  \bibinfo {author} {\bibfnamefont {N.}~\bibnamefont {Rivas}}, \bibinfo
  {author} {\bibfnamefont {A.}~\bibnamefont {Scagliarini}}, \bibinfo {author}
  {\bibfnamefont {J.}~\bibnamefont {Harting}}, \ and\ \bibinfo {author}
  {\bibfnamefont {S.}~\bibnamefont {Dietrich}},\ }\href@noop {} {\bibfield
  {journal} {\bibinfo  {journal} {Soft Matter}\ }\textbf {\bibinfo {volume}
  {16}},\ \bibinfo {pages} {3536} (\bibinfo {year} {2020})}\BibitemShut
  {NoStop}%
\bibitem [{\citenamefont {Michelin}\ and\ \citenamefont
  {Lauga}(2014)}]{Michelin2014}%
  \BibitemOpen
  \bibfield  {author} {\bibinfo {author} {\bibfnamefont {S.}~\bibnamefont
  {Michelin}}\ and\ \bibinfo {author} {\bibfnamefont {E.}~\bibnamefont
  {Lauga}},\ }\href {\doibase 10.1017/jfm.2014.158} {\bibfield  {journal}
  {\bibinfo  {journal} {J. Fluid Mech.}\ }\textbf {\bibinfo {volume} {747}},\
  \bibinfo {pages} {572} (\bibinfo {year} {2014})}\BibitemShut {NoStop}%
\bibitem [{\citenamefont {Michelin}\ and\ \citenamefont
  {Lauga}(2015)}]{Michelin2015}%
  \BibitemOpen
  \bibfield  {author} {\bibinfo {author} {\bibfnamefont {S.}~\bibnamefont
  {Michelin}}\ and\ \bibinfo {author} {\bibfnamefont {E.}~\bibnamefont
  {Lauga}},\ }\href@noop {} {\bibfield  {journal} {\bibinfo  {journal} {Eur.
  Phys. J. E}\ }\textbf {\bibinfo {volume} {38}},\ \bibinfo {pages} {1}
  (\bibinfo {year} {2015})}\BibitemShut {NoStop}%
\bibitem [{\citenamefont {Strogatz}(2015)}]{Strogatz2015}%
  \BibitemOpen
  \bibfield  {author} {\bibinfo {author} {\bibfnamefont {S.}~\bibnamefont
  {Strogatz}},\ }\href@noop {} {\emph {\bibinfo {title} {Nonlinear Dynamics and
  Chaos}}}\ (\bibinfo  {publisher} {CRC Press},\ \bibinfo {address} {Boca
  Raton},\ \bibinfo {year} {2015})\BibitemShut {NoStop}%
\bibitem [{\citenamefont {Zwanzig}(1992)}]{Zwanzig1992}%
  \BibitemOpen
  \bibfield  {author} {\bibinfo {author} {\bibfnamefont {R.}~\bibnamefont
  {Zwanzig}},\ }\href@noop {} {\bibfield  {journal} {\bibinfo  {journal} {J.
  Phys. Chem.}\ }\textbf {\bibinfo {volume} {96}},\ \bibinfo {pages} {3926}
  (\bibinfo {year} {1992})}\BibitemShut {NoStop}%
\bibitem [{\citenamefont {Reguera}\ and\ \citenamefont
  {Rubi}(2001)}]{Reguera2001}%
  \BibitemOpen
  \bibfield  {author} {\bibinfo {author} {\bibfnamefont {D.}~\bibnamefont
  {Reguera}}\ and\ \bibinfo {author} {\bibfnamefont {J.~M.}\ \bibnamefont
  {Rubi}},\ }\href@noop {} {\bibfield  {journal} {\bibinfo  {journal} {Phys.
  Rev. E}\ }\textbf {\bibinfo {volume} {64}},\ \bibinfo {pages} {061106}
  (\bibinfo {year} {2001})}\BibitemShut {NoStop}%
\bibitem [{\citenamefont {Malgaretti}\ \emph {et~al.}(2013)\citenamefont
  {Malgaretti}, \citenamefont {Pagonabarraga},\ and\ \citenamefont
  {Rubi}}]{Malgaretti2013}%
  \BibitemOpen
  \bibfield  {author} {\bibinfo {author} {\bibfnamefont {P.}~\bibnamefont
  {Malgaretti}}, \bibinfo {author} {\bibfnamefont {I.}~\bibnamefont
  {Pagonabarraga}}, \ and\ \bibinfo {author} {\bibfnamefont {J.}~\bibnamefont
  {Rubi}},\ }\href@noop {} {\bibfield  {journal} {\bibinfo  {journal} {Front.
  Phys.}\ }\textbf {\bibinfo {volume} {1}},\ \bibinfo {pages} {21} (\bibinfo
  {year} {2013})}\BibitemShut {NoStop}%
\bibitem [{\citenamefont {Ebbens}\ \emph {et~al.}(2012)\citenamefont {Ebbens},
  \citenamefont {Tu}, \citenamefont {Howse},\ and\ \citenamefont
  {Golestanian}}]{Ebbens2012}%
  \BibitemOpen
  \bibfield  {author} {\bibinfo {author} {\bibfnamefont {S.}~\bibnamefont
  {Ebbens}}, \bibinfo {author} {\bibfnamefont {M.-H.}\ \bibnamefont {Tu}},
  \bibinfo {author} {\bibfnamefont {J.~R.}\ \bibnamefont {Howse}}, \ and\
  \bibinfo {author} {\bibfnamefont {R.}~\bibnamefont {Golestanian}},\ }\href
  {\doibase 10.1103/PhysRevE.85.020401} {\bibfield  {journal} {\bibinfo
  {journal} {Phys. Rev. E}\ }\textbf {\bibinfo {volume} {85}},\ \bibinfo
  {pages} {020401} (\bibinfo {year} {2012})}\BibitemShut {NoStop}%
\bibitem [{\citenamefont {Kim}\ \emph {et~al.}(2013)\citenamefont {Kim},
  \citenamefont {Yokokawa},\ and\ \citenamefont {Takayama}}]{Kim2013}%
  \BibitemOpen
  \bibfield  {author} {\bibinfo {author} {\bibfnamefont {S.-J.}\ \bibnamefont
  {Kim}}, \bibinfo {author} {\bibfnamefont {R.}~\bibnamefont {Yokokawa}}, \
  and\ \bibinfo {author} {\bibfnamefont {S.}~\bibnamefont {Takayama}},\
  }\href@noop {} {\bibfield  {journal} {\bibinfo  {journal} {Lab Chip}\
  }\textbf {\bibinfo {volume} {13}},\ \bibinfo {pages} {1644} (\bibinfo {year}
  {2013})}\BibitemShut {NoStop}%
\bibitem [{\citenamefont {Palacci}\ \emph {et~al.}(2014)\citenamefont
  {Palacci}, \citenamefont {Sacanna}, \citenamefont {Kim}, \citenamefont {Yi},
  \citenamefont {Pine},\ and\ \citenamefont {Chaikin}}]{Palacci2014}%
  \BibitemOpen
  \bibfield  {author} {\bibinfo {author} {\bibfnamefont {J.}~\bibnamefont
  {Palacci}}, \bibinfo {author} {\bibfnamefont {S.}~\bibnamefont {Sacanna}},
  \bibinfo {author} {\bibfnamefont {S.-H.}\ \bibnamefont {Kim}}, \bibinfo
  {author} {\bibfnamefont {G.-R.}\ \bibnamefont {Yi}}, \bibinfo {author}
  {\bibfnamefont {D.~J.}\ \bibnamefont {Pine}}, \ and\ \bibinfo {author}
  {\bibfnamefont {P.~M.}\ \bibnamefont {Chaikin}},\ }\href@noop {} {\bibfield
  {journal} {\bibinfo  {journal} {Philosophical Transactions of the Royal
  Society A: Mathematical, Physical and Engineering Sciences}\ }\textbf
  {\bibinfo {volume} {372}},\ \bibinfo {pages} {20130372} (\bibinfo {year}
  {2014})}\BibitemShut {NoStop}%
\bibitem [{\citenamefont {Schlichting}(1979)}]{Schlichting1979}%
  \BibitemOpen
  \bibfield  {author} {\bibinfo {author} {\bibfnamefont {H.}~\bibnamefont
  {Schlichting}},\ }\href@noop {} {\emph {\bibinfo {title} {Boundary Layer
  Theory}}}\ (\bibinfo  {publisher} {McGraw-Hill},\ \bibinfo {address} {New
  York},\ \bibinfo {year} {1979})\BibitemShut {NoStop}%
\end{thebibliography}%

\clearpage

\onecolumngrid
\setcounter{equation}{0}
\setcounter{figure}{0}
\renewcommand\theequation{S\arabic{equation}}
\renewcommand\thefigure{S\arabic{figure}}
\section*{Supplementary Material}

\section{Coupling of the Navier-Stokes equation with the solute density}
The presence of the interaction potential $U_{wall}$ between the wall and the solute molecules leads to a body force term in the Navier-Stokes equation:
\begin{equation}
\label{eq:navierstokes}
\rho_f \left[ \frac{\partial}{\partial t} \bm{v} + (\bm{v} \cdot \nabla) \bm{v} \right] = - \nabla P + \mu \nabla^2 \bm{v} - \rho \nabla U_{wall}, 
\end{equation}
where $\rho_f$ is the fluid mass density, $P$ is the pressure, and $\mu$ is the dynamic viscosity. This body force drives the hydrodynamics. Furthermore, the continuity equation applies:
\begin{equation}
\label{eq:continuity}
\frac{\partial}{\partial t} \rho_f = - \nabla (\rho_f \bm{v}).
\end{equation}
In the present work, we have focused on the limit of low Reynolds and Mach numbers for Eqs.~\eqref{eq:navierstokes} and~\eqref{eq:continuity}. For high values of $Pe$ and of $a_{cov}$, the details of the initial condition may determine whether the pore fails to pump ($\tilde{Q}=0$), pumps steadily, or displays persistent oscillations (see overlapping crosses and  open/solid points in Fig.~\ref{fig:diagram}). The occurrence of two possible asymptotic dynamics for the same pore is due to the term $\rho\bm{v}[\rho]$ in Eq.~\eqref{eq:rho} which is non-linear in $\rho$ and couples to Eq.~\eqref{eq:navierstokes}. However, a complete study of the basins of attraction of the diverse steady-states is beyond the scope of the current study.

\section{Definition of the power spectrum}
The power spectrum $S_{\tilde{Q}\tilde{Q}}$ is defined as
\begin{equation}
    S_{\tilde{Q}\tilde{Q}} (\omega) = \left|\frac{1}{2T_{rel}}\int_{T_{rel}}^{3T_{rel}} \tilde{Q}(t) e^{-2\pi i \omega t} \ dt \right|^2,
\end{equation} 
where $T_{rel}$ is the relaxation time beyond which we consider the simulations to have attained sustained oscillations. For all simulations, we have $T_{rel}\tau_f^{-1}=1.74$.

\section{Definition of the P\'eclet number}

\begin{figure*}[ht]
	\centering
   \includegraphics[width=0.5\textwidth]{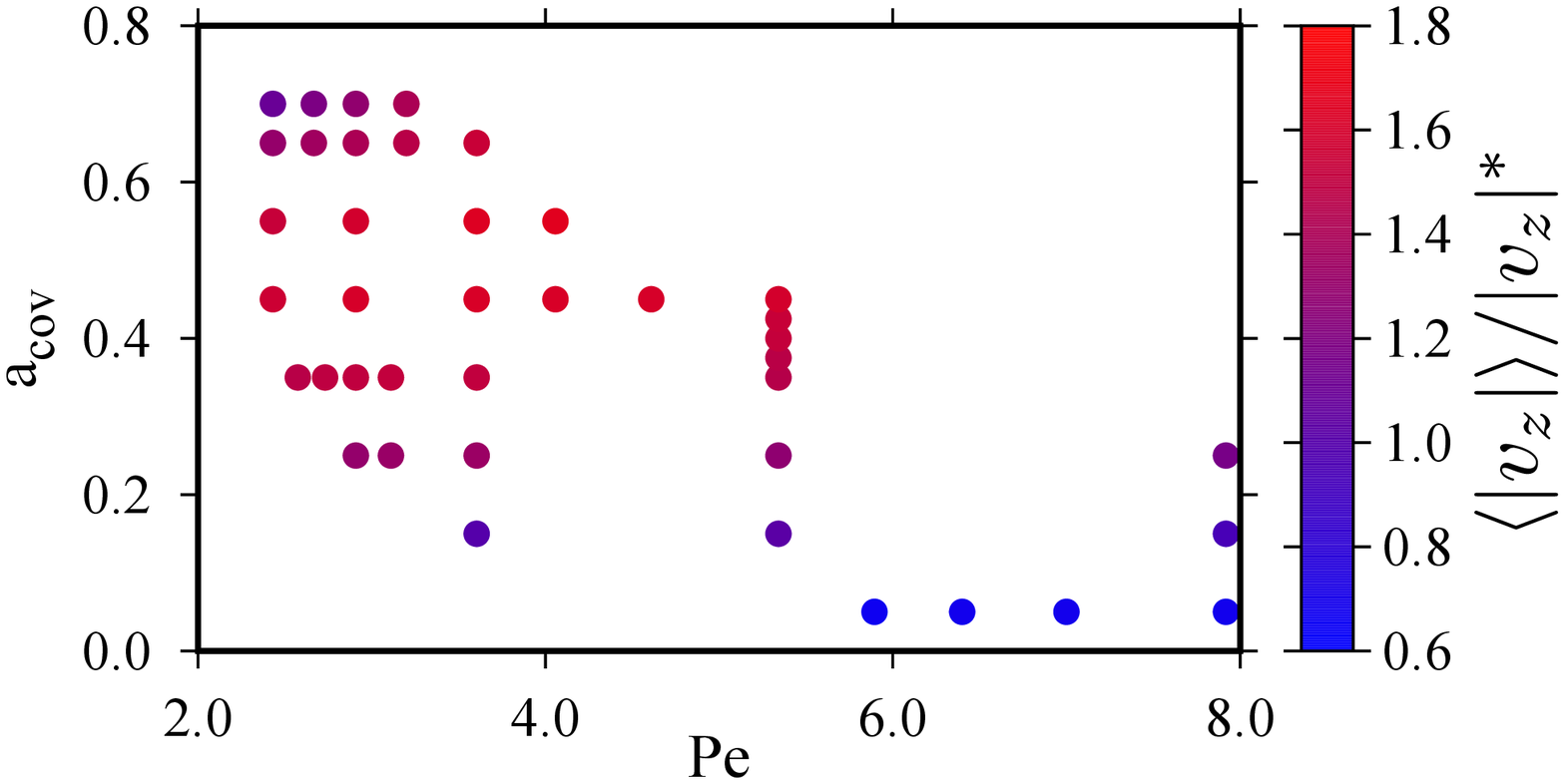}
   \caption{ Normalized average velocity $\langle |v_z| \rangle$ (Eq.~\eqref{eq:v*}) near the wall as function of $Pe$ and $a_{cov}$. The normalization is taken with $|v_z|^* = (2L)/\tau_f$, where $\tau_f = (2L)^2/\nu$ is the relaxation time of the fluid. In lattice units, we have chosen the parameters $L=20$, $\nu = 1/6$, $U_0 = 4 \times 10^ {-4}$, $l=4$, $\xi=1$, $\chi=10^{-3}$, $\beta=1$, and $\theta = \pi/6$. The simulation box encompasses $80\times80\times40$ lattice units.}
    	\label{fig:vel}
\end{figure*}

The P\'eclet number
\begin{equation}
\label{eq:peclet}
Pe = \frac{v^*L}{D}
\end{equation} 
requires to introduce a characteristic velocity scale $v^*$. We determine $v^*$ from the average velocity of the fluid in a shell around the wall, as it is the solute in this region which drives diffusioosmosis. This shell comprises the region in which the potential $U_{wall}$ is non-zero, and thus it is narrow compared with the pore radius. We carry out these measurements at steady pumping states. Pumping is maintained via advection in the solute-poor half of the pore and by carrying solute to the solute-rich half, where it eventually leaves this segment of the pore. Since $v^*$ characterizes advection, the integration for taking the average is performed only in the solute-poor half. The average of the absolute value of the z-component of the velocity near the wall is
\begin{equation}
	\label{eq:v*}
    \langle |v_z| \rangle = S\int_{-L}^0 dz \int_{R(z)-l}^{R(z)} dr \ r\int_0^{2\pi} d\phi \  |v_z(r,\phi,z)|,
\end{equation}
where
\begin{equation}
        S = \left[ \int_{-L}^0 dz  \int_{R(z)-l}^{R(z)} dr \ r \int_0^{2\pi} d \phi \right]^{-1},
\end{equation}
 with the results shown in Fig.~\ref{fig:vel}. The velocity $\langle |v_z| \rangle$ is a function of $D$ and $a_{cov}$, with the maximum value being at most a factor of three larger than the minimum value. In order to capture the common order of magnitude, we define $v^*$ as an average over the phase space:
\begin{equation}
 v^* = S^{\prime}\iint\limits_{\Omega_{p}} dD da_{cov}  \  \langle |v_z| \rangle(a_{cov},D),
\end{equation} 
where
\begin{equation}
 S^{\prime}=\left[\iint\limits_{\Omega_{p}} dD da_{cov}\right]^{-1},
\end{equation} 
and $\Omega_{p}$ is the region of phase space exhibiting steady pumping which we have sampled. This procedure renders $v^*=1.3 |v_z|^*$ ($0.0056$ in lattice units). This value is used throughout the text. 

\section{Fick-Jacobs-based theory}

\subsection{Analytical approach}

\begin{figure}[t]
	\centering
   \includegraphics[width=1.0\textwidth]{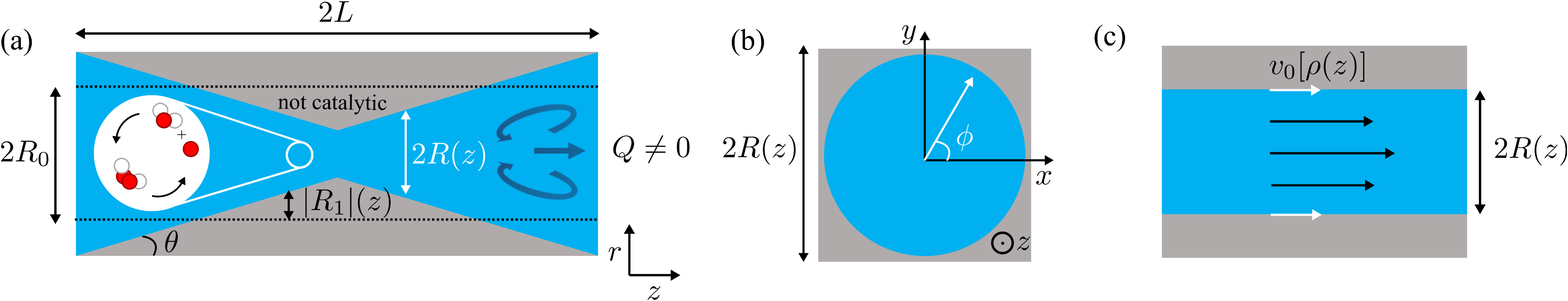}
   \caption{(a) Cartoon of the longitudinal cross-section of the active pore. The pore length is $2L$ and its variable radius is $R(z)$, with an average radius of $R_0$. The catalysis of the chemical solute occurs in the bulk (blue region) with a possibly spatially inhomogeneous rate. The reverse chemical reaction destroys solute also in the bulk with a spatially homogeneous rate (see inset). The pore walls interact with the solution via the effective potential $U_{wall}$. Convection rolls (dark blue arrows) emerge due to diffusioosmosis and eventually may lead to the onset of a net non-zero flow rate $Q$. (b) Transverse cross-section of the active pore. (c) Velocity profile induced by diffusioosmosis in a flat section of the pore. The velocity at the wall is a functional of the number density of solute and is given by, c.f., Eq.~\eqref{eq:vslip}.
   }
    	\label{fig:modelNotes}
\end{figure}

In this section, we show how the spontaneous symmetry breaking observed in the simulations can be revealed also semi-analytically. To this end, we employ a Fick-Jacobs-based technique in order to reduce the dimensionality of the description~\cite{Zwanzig1992,Reguera2001,Malgaretti2013}. We consider an axially symmetric pore with its axis along the z direction. The cross-section is thus circular with a radius which varies with z. The ends of the pore are located at $z = \pm L$. The pore walls are located at $x^2 + y^2=R^2(z)$, with the average radius $R_0$ defined as 
\begin{equation}
    R_0 = \frac{1}{2L}\int\limits_{-L}^L R(z) dz .
\end{equation}
See Fig.~\ref{fig:modelNotes} (a) for a sketch of the pore geometry. In the following we assume that a chemical reaction $A\rightarrow B$ occurs in the bulk with the rate $\hat{\xi}_B(\mathbf{r})$, and that the reverse reaction $B\rightarrow A$ occurs also in the bulk with the rate $\chi \rho(\mathbf{r},t)$, where $\rho$ is the number density of the reaction products (called solute from here on). For the sake of simplicity, we fix the number density of the reactant (species $A$) to be homogeneous in space and constant in time. 

On top of the chemical reaction, the solute number density changes due to diffusion, due to advection, and due to the interaction with the pore walls. Thus, the solute number density $\rho$ is governed by the differential equation
\begin{equation}
\dot{\rho}(\mathbf{r},t)=-\nabla\cdot \textbf{j}(\mathbf{r},t)+\hat{\xi}_B(\mathbf{r})-\chi\rho(\mathbf{r},t)\,,\label{eq:adv-diff}
\end{equation}
where $\hat{\xi}_B(\mathbf{r})$ is a non-negative source term, $\chi$ is a positive sink constant, and the flux $\textbf{j}(\mathbf{r},t)$ is given by
\begin{equation}
\textbf{j}(\mathbf{r},t) = -D\nabla\rho(\mathbf{r},t) -\beta D\rho(\mathbf{r},t)\nabla W(\mathbf{r}) + \bm{v}(\mathbf{r},t)\rho(\mathbf{r},t);
\label{eq:app-J-0}
\end{equation}
$\bm{v}$ is the velocity profile, and
\begin{equation}
W(\mathbf{r})=\begin{cases}
U_{wall}(\mathbf{r}), & |\mathbf{r}| \leq R(z)\,,\\
\infty, & \text{otherwise},
\end{cases}\label{eq:pot}
\end{equation}
encodes the Hamiltonian interaction $U_{wall}(\mathbf{r})$ with the pore walls and the confinement of the solute inside the pore. Here we are interested in the regime in which the advection due to the longitudinal fluid flow (i.e., the contribution $v_z\rho$ in Eq.~\eqref{eq:app-J-0}) dominates the term $\beta D\rho(\mathbf{r},t)\partial_z W(\mathbf{r})\textbf{e}_z$ in Eq.~\eqref{eq:app-J-0}. This is particularly valid if the pore radius varies weakly, i.e., $\partial_z R(z)\ll 1$. Accordingly, Eq.~\eqref{eq:app-J-0} reduces to 
\begin{equation}
\textbf{j}(\mathbf{r},t) = -D\nabla\rho(\mathbf{r},t) -\beta D\rho(\mathbf{r},t)\partial_x W(\mathbf{r}) \textbf{e}_x -\beta D\rho(\mathbf{r},t)\partial_y W(\mathbf{r}) \textbf{e}_y + \bm{v}(\mathbf{r},t)\rho(\mathbf{r},t)\,,
\label{eq:app-J}
\end{equation}
where $\textbf{e}_x$ and $\textbf{e}_y$ are the unit vectors along the x and the y axis, respectively. Due to the cylindrical symmetry of the pore, it is convenient to switch to cylindrical coordinates with $r$ as the distance from the $z$ axis and $\phi$ as the angle of \textbf{r} formed with the xz-plane (Fig.~\ref{fig:modelNotes} (b)). 
Integrating Eq.~\eqref{eq:adv-diff} over the cross-section and exploiting the axial symmetry of the system leads to
\begin{align}
\int_{0}^{\infty} \dot{\rho}(r,z) r dr = &- \int_{0}^{\infty} \partial_z j_z(r,z) rdr - \int_{0}^{\infty} \partial_r (r j_r(r,z)) dr +  \int_{0}^{\infty}(\hat{\xi}_B(r,z)-\chi\rho)rdr\nonumber\\
&=- \int_{0}^{\infty} \partial_z j_z(r) rdr +  \int_{0}^{\infty}(\hat{\xi}_B(r,z)-\chi\rho) r dr.
\label{eq:adv-diff2}
\end{align}

The above step exploits the fact that, due to Eq.~\eqref{eq:pot}, there is no solute anywhere outside the pore, i.e.,
\begin{align}
    \rho \Big(r > R(z)\Big)=0,
\end{align}
and
\begin{align}
    j_r\Big(r > R(z)\Big)=0\,.
\end{align}
Therefore, the integrals over $r$ in Eq.~\eqref{eq:adv-diff2} eventually reduce to integrals inside the pore only. However, keeping the upper limit as infinity greatly facilitates the calculation. The incompressibility equation applies, i.e.,
\begin{align}
    \nabla\cdot\bm{v}(\mathbf{r},t)=\frac{1}{r}\partial_r\left(r v_r(\mathbf{r},t)\right)+\partial_z v_z(\mathbf{r},t)= 0\,,
    \label{eq:app-incompr}
\end{align}
where in the last step we have assumed that the axial symmetry of the pore leads to an axial symmetry of the velocity profile, i.e., $\partial_\phi \bm{v}=0$.
Equation~\eqref{eq:app-incompr} allows us to define the order of magnitude of the radial component $v^*_r$ of the velocity as
\begin{equation}
\label{eq:radialv}
    v^*_r = \frac{R_0}{L} v^*_z,
\end{equation}
where $v^*_z$ is the order of magnitude of the longitudinal component of the velocity (as per Eq.~\eqref{eq:v*}):
\begin{equation}
	\label{eq:v*_2}
    v_z^* = 2 \pi S \int_{-L}^0 dz \ R(z) \sqrt{1 + \partial_zR(z)^2} \ |v_z(r=R(z),z)|,
\end{equation}
where
\begin{equation}
        S = \left[ 2 \pi \int_{-L}^0 dz \ R(z) \sqrt{1 + \partial_zR(z)^2} \right]^{-1}.
\end{equation}
We introduce the radial P\'eclet number $Pe_r$ by quantifying the relative magnitude of diffusive and advective timescales in the radial direction:
\begin{equation}
    Pe_r = \frac{R_0 v^*_r}{D},
\end{equation}
and as per Eq.~\eqref{eq:radialv}
\begin{align}
    Pe_r= \frac{R^2_0}{L}\frac{v_z^*}{D}.
\end{align}
In a sufficiently narrow pore ($R_0 \ll L)$, one has $Pe_r \ll 1$. This means that diffusive and potential-driven transport dominates the advection due to the fluid flow in the radial direction. Furthermore --- for the same diffusion coefficient and the same potential --- the shorter the distance a solute needs to be transported in order to obtain equilibration, the shorter the associated timescale. Thus, in a narrow pore, the relaxation of $\rho$ along the radial direction occurs on shorter time scales as compared to the longitudinal direction. Accordingly, we assume no net transport along the radial direction (Fick-Jacobs approximation~\cite{Zwanzig1992,Reguera2001,Malgaretti2013}), or
\begin{equation}
\label{eq:fickjacobs}
   \partial_r (rj_r) = 0.
\end{equation}
In order to gain analytical insight, in the following we assume that the source term is constant along the radial direction:
\begin{equation}
\hat{\xi}_B(z)=\begin{cases}
\frac{\xi_B(z)}{\pi R^2(z)} & ,r<R(z)\,,\\
0 & \text{, otherwise.}\,
\end{cases}\label{eq:source}
\end{equation}
Such an assumption is at odds with the numerical simulations which are based on a chemical reaction which occurs \textit{solely at the pore walls}. Since the Fick-Jacobs  approximation is crucial for obtaining a model which can be analytically solved, we shall compare the results stemming from that model with those results which belong to the numerical simulations. Indeed, the good agreement between these numerical data and the predictions of the model justify this approximation \textit{a posteriori}.
In the absence of fluxes through the pore walls, Eqs.~ \eqref{eq:fickjacobs} and \eqref{eq:source} lead to the following \textit{ansatz} (Fick-Jacobs approximation):
\begin{equation}
\rho(r,z,t)  =p(z,t)\frac{1}{R^2_{0}}\frac{e^{-\beta W(r,z)}}{e^{-\beta A(z)}},\label{eq:rho-ansatz}
\end{equation}
with $W(r,z)$ given by Eq.~\eqref{eq:pot} and
\begin{equation}
 R_0^2e^{-\beta A(z)} =2 \pi\int_{0}^{\infty} e^{-\beta W(r,z)}rdr\,.
\end{equation}
Accordingly, inserting Eq.~\eqref{eq:rho-ansatz} into Eq.~\eqref{eq:adv-diff2} and using Eq.~\eqref{eq:app-J} leads to
\begin{equation}
\dot{p}(z,t) = \partial_{z}\left[D\partial_{z}p(z,t)-\frac{Q(z,t)}{e^{-\beta A(z)}}\frac{p(z,t)}{R_0^2}\right]+\xi_B(z)-\chi p(z,t) \label{eq:FJ1}
\end{equation}
with
\begin{equation}
Q(z,t) \equiv 2 \pi \int_{0}^{\infty} v_z(r,z,t) e^{-\beta W(r,z)} rdr\label{eq:def_Q}\,.
\end{equation}
The range $l$ (see Eq.~\eqref{eq:potentialMain}) of the interactions between the solute molecules and the pore walls is typically of molecular size. Hence, in order to keep the model as simple as possible, in the following we assume that $l\ll R_0$, so that $e^{-\beta W} \approx 1$ for $r < R(z)$, and $e^{-\beta W} \ll 1$ for $r > R(z)$. Since the dependence of $\rho$ on $r$ enters only via $\exp(-\beta W)$, one concludes that $\rho$ does not depend on $r$. This assumption allows one to simplify Eq.~\eqref{eq:rho-ansatz}:
\begin{equation}
\label{eq:ansatzIsotropic}
\rho(r,z,t) \equiv \rho(z) \ \simeq \ \frac{p(z,t)}{\pi R^2(z)}\,,
\end{equation}
which implies
\begin{equation}
\dot{p}(z,t)=D\partial_z^2p(z,t)  - \partial_z\Big[Q(t)\frac{p(z,t)}{\pi R^2(z)}\Big] + \xi_B(z) - \chi p(z,t),\label{eq:toExpand}
\end{equation}
with
\begin{equation}
Q(t) = 2 \pi \int_{0}^{R(z)} v_z(r,z,t)r dr.
\label{eq:Q_comp}
\end{equation}
We note that Eq.~\eqref{eq:Q_comp} is the fluid flow rate which, due to the incompressibility (see Eq.~\eqref{eq:app-incompr}), does not vary along the $z$ direction. We note that Eq.~\eqref{eq:def_Q} can be reduced to Eq.~\eqref{eq:Q_comp} due to the approximation in Eq.~\eqref{eq:source} which eventually leads to Eq.~\eqref{eq:ansatzIsotropic}. 

Finally, at steady state, Eqs.~\eqref{eq:toExpand}, and \eqref{eq:Q_comp} reduce to 
\begin{equation}
0=D\partial_z^2p(z)  - \partial_z\Big[Q\frac{p(z)}{\pi R^2(z)}\Big] + \xi_B(z) - \chi p(z),\label{eq:toExpand_st}
\end{equation}
with
\begin{equation}
Q= 2 \pi \int_{0}^{R(z)} v_z(r,z)r dr, \label{eq:Q_comp_st}
\end{equation}
where, in order to keep the notation simple, we mark the steady state solution by omitting the time dependence. 

In order to solve the problem, one finally has to express $Q$ in terms of $p(z)$. Indeed, the inhomogeneous distribution of solute induces a local phoretic slip velocity relative to the stationary pore walls. The z-component of this velocity is \cite{Anderson1989}.
\begin{equation}
\label{eq:vslip}
v_{0}(z)=-\frac{\mathcal{L}}{\beta\eta}\nabla_{||}\rho(r,z) \cdot \textbf{e}_z\,,
\end{equation}
where $\mathcal{L}$ is the phoretic mobility (a negative constant which stems from a repulsive potential between the wall and the solute molecules), $\nabla_{||}$ is the derivative along the surface evaluated at the wall, and $\textbf{e}_z$ is the unit vector in the z direction. Due to the small scales of the pore ($Re \ll 1$), we describe the fluid via the Stokes equation together with the condition for incompressibility:
\begin{equation}\label{eq:app-stokes1}
\eta\nabla^{2}\bm{v}(r,z)      = \nabla P(r,z),
\end{equation}
and
\begin{equation}\label{eq:app-stokes2}
\nabla\cdot \bm{v}(r,z) = 0,
\end{equation}
where $\eta$ is the dynamic viscosity, $\bm{v}$ is the flow velocity, and $P$ is the pressure. Equation~\eqref{eq:vslip} acts as a boundary condition for the Stokes equation. As mentioned early on, the pore is considered to be narrow ($L \gg R_0$) and axially symmetric, as is the velocity profile. Within such a regime, we exploit the lubrication approximation \cite{Schlichting1979} ($\partial_z^2 v_z \ll r^{-1} \partial_r (r \partial_r v_z)$) and solve Eqs.~\eqref{eq:app-stokes1} and~\eqref{eq:app-stokes2}. This leads to
\begin{equation}
v_z(r,z) = v_0(z) - \frac{\partial_z P(z)}{4 \eta} \Big[ R^2(z) - r^2 \Big]\,,
\end{equation}
where $v_0(z) = v_z(r=R,z)$. Note that $\partial_{z}P(z)$ contains, in addition to a possible external pressure drop, also contributions from the Lagrange multiplier implementing $\nabla\cdot \bm{v}=0$. As already mentioned, fluid incompressibility corresponds to a fluid flow rate $Q$ which is constant in time:
\begin{align}
Q=2 \pi \int_{0}^{R(z)} v_z(r,z) rdr=v_0(z) \pi R^2(z)-\frac{\pi}{8}\frac{\partial_z P(z)}{ \eta}R^4(z)\,.
\end{align}
In the following we consider the special case in which there is no external pressure drop. Accordingly, $P(z)$ has to fulfill periodic boundary conditions, i.e, the integral of $\partial_z P(z)$ over the pore length must vanish:
\begin{align}
    0=\int_{-L}^L \partial_z P(z) dz = 8 \pi \int_{-L}^L \frac{v_0(z)}{R^{2}(z)} dz - \frac{8 \eta}{\pi}Q \int_{-L}^L\frac{dz}{R^{4}(z)}.
\end{align}
This allows one to determine $Q$:
\begin{equation}
Q  = \pi \int\limits_{-L}^ L \dfrac{v_0(z)}{R^2(z)} dz  \Big/  \int\limits_{-L}^ L \dfrac{dz}{R^4(z)} \,.
\label{eq:Q}
\end{equation}
In the next step we obtain the slip velocity $v_0(z)$ as a function of $p(z)$. The vector perpendicular to the pore wall is $\textbf{n}(\phi,z)$:
\begin{equation}
    \textbf{n}(\phi,z) = \frac{1}{\sqrt{1 + \Big(\partial_z R(z)\Big)^2}}[- \textbf{e}_r + \partial_z R(z) \textbf{e}_z]\,,
\end{equation}
where $\textbf{e}_r$ is the unit vector pointing in the radial direction. With this Eq.~\eqref{eq:vslip} turns into
\begin{align}
v_{0}(z) &= -\frac{\mathcal{L}}{\beta\eta} \Big[ \nabla\rho(r=R) \cdot \bf{e}_z - (\nabla\rho(r=R) \cdot \bf{n}) \bf{n} \cdot \bf{e}_z \Big] =  \nonumber \\
		&=-\frac{\mathcal{L}}{\beta\eta}\Big[ \partial_z\rho-\frac{(\partial_z \rho) (\partial_z R) + r^{-1}\partial_r(r\partial_r \rho)}{\sqrt{1 + (\partial_z R)^2}} \frac{\partial_z R}{\sqrt{1 + (\partial_z R)^2}}\Big]\,, \label{eq:v0toapproximate}
\end{align}
and, with Eq.~\eqref{eq:ansatzIsotropic}, $\partial_r \rho =0$. We recall that Eq.~\eqref{eq:app-J} is particularly valid in the regime of weakly varying pore radii. In this regime one has $ (\partial_z R)^2 \ll 1 $, and Eq.~\eqref{eq:v0toapproximate} can be approximated as
\begin{align}
\label{eq:v02}
v_{0}(z) \approx -\frac{\mathcal{L}}{\beta\eta}\partial_z\rho (z,r=R)  \ =  -\frac{\mathcal{L}}{\beta\eta}\partial_z\left[\frac{p(z)}{\pi R^2(z)} \right].
\end{align}
In order to gain analytic insight we expand Eq.~\eqref{eq:toExpand_st} around $Q=0$. Accordingly, we expand $p(z)\equiv p(z,Q)$:
\begin{equation}
\label{eq:expansionQ}
p(z;Q) = p_0(z) + \sum\limits_{j>0} p_j(z)Q^j.
\end{equation}
The functions $p_0(z)$ and $p_j(z)$ are independent of $Q$. Plugging Eq.~\eqref{eq:expansionQ} into Eq.~\eqref{eq:toExpand_st} leads to
\begin{align}
\label{eq:eqAfterExpansion}
0 = & D \partial_z^2 \left(p _0(z) + \sum\limits_{j>0} p_j(z)Q^j \right) + \xi_B(z)  - \chi \left(p _0(z) + \sum\limits_{j>0} p_j(z)Q^j \right)  \\
& - \frac{Q}{\pi}  \partial_z \left[ \left(p _0(z) + \sum\limits_{j>0} p_j(z)Q^j \right)  R^{-2}(z) \right].
\end{align} 
Grouping the terms together in accordance with the same order of $Q$, and by using the fact that Eq.~\eqref{eq:eqAfterExpansion} must be valid for any value of $Q$, we obtain a hierarchy of equations, starting from an equation for the zeroth order contribution:
\begin{equation}
0= D \partial_z^2 p_0(z) + \xi_B(z) - \chi p_0(z), \label{eq:expansion0}
\end{equation}
the solution of which is then used to recursively determine the higher order contributions:
\begin{equation}
0= D \partial_z^2 p_j(z) - \chi p_j(z) - \frac{1}{\pi} \partial_z \Big[ R(z)^{-2} p_{j-1}(z) \Big]  \text{ , } j > 0.\label{eq:expansion1}
\end{equation}
By using Eqs.~\eqref{eq:ansatzIsotropic},~\eqref{eq:v02}, and Eq.~\eqref{eq:expansionQ}, Eq.~\eqref{eq:Q} reads
\begin{equation}
\label{eq:Q_expanded}
Q  = -\frac{\mathcal{L}}{\beta \eta} \left[ \int\limits_{-L}^ L \dfrac{dz}{R^4(z)} \right]^{-1} \int\limits_{-L}^ L R^{-2}(z) \left\{\partial_z[ R^{-2}(z) p_0(z) ]  + \sum\limits_{j > 0} Q^j\partial_z[ R^{-2}(z) p_j(z) ]  \right\} dz.
\end{equation}
We proceed by representing the pore radius $R(z)$ and the volumetric source strength $\hat{\xi}_B(z) = \pi^{-1} R(z)^{-2} \xi_B(z)$ in terms of their Fourier coefficients:
\begin{equation}
\label{eq:defR}
R^{-2}(z)= \alpha_0 + \SUMIGRZERO \alpha_i \cos(k_i z)
\end{equation}
and
\begin{equation}
\label{eq:defXi}
\xi_B(z)= \xi_{B,0} + \SUMIGRZERO \xi_{B,i} \cos(k_i z), \ k_i = \frac{\pi}{L}i.
\end{equation}
The quantities $\alpha_0$ and $\xi_{B,0}$ correspond to the mean values 
\begin{equation}
\alpha_0 = \frac{1}{2L}\int\limits_{-L}^L R^{-2}(z)dz
\end{equation} 
and
\begin{equation}
\xi_{B,0} = \frac{1}{2L}\int\limits_{-L}^L \xi_B(z)dz,
\end{equation} 
respectively, while the coefficients $\alpha_i$ and $\xi_{B,i}$ control the variations around these means. These variations are left-right symmetric (i.e., w.r.t $z \leftrightarrow -z$) and later they will be chosen as to match the Lattice Boltzmann simulations. The mapping between the bulk source $\hat{\xi}_B(z)$ and the surface source employed in the simulations will be specified later. We proceed by performing the spatial Fourier expansions of $p_0(z)$ and $p_i(z)$ in space:
\begin{equation}
\label{eq:fourierp0}
p_0(z) = p_{0,0} + \sum\limits_{i>0} p_{0,i} \cos(k_i z) + \sum\limits_{i>0} \tilde{p}_{0,i} \sin(k_i z),
\end{equation}
and
\begin{equation}
p_j(z)= p_{j,0} + \sum\limits_{i>0} p_{j,i} \cos(k_i z) + \sum\limits_{i>0} \tilde{p}_{j,i} \sin(k_i z)   \text{ , } j > 0.
\label{eq:fourierpj}
\end{equation}
Plugging these two expansions (Eqs.~\eqref{eq:fourierp0} and \eqref{eq:fourierpj}), as well as the definitions for $R^{{-2}}(z)$ and $\xi_B(z)$ (Eqs.~\eqref{eq:defR} and \eqref{eq:defXi}) into Eqs. \eqref{eq:expansion0} and \eqref{eq:expansion1}, leads to closed formulae for the functions $p_j(z)$ for any desired value of $j$. The contributions of order zero in $Q$ and $p_0(z)$ are obtained by solving Eq.~\eqref{eq:expansion0} together with Eq. \eqref{eq:defXi}, in order to obtain the Fourier coefficients
\begin{equation}
p_{0,0}= \frac{\xi_{B,0}}{\chi} \label{eq:p00} 
\end{equation}
and
\begin{equation}
p_{0,i} = \frac{\xi_{B,i}}{\chi + D k_i^2}, \label{eq:p0r}
\end{equation}
and finally for $z \leftrightarrow -z$ symmetry reasons  (Eq.~\eqref{eq:expansion0})
\begin{equation}
\tilde{p}_{0,i}= 0 \text{ , for } i > 0,
\end{equation}
so that $p_0(z)$ is an even function. Therefore, and due to $R(z)=R(-z)$, the first term in the curly brackets in Eq.~\eqref{eq:Q_expanded} does not contribute to that integral.
Indeed, the even modes of all functions $p_j(z)$ do not contribute to the integral. While the hierarchy of equations given by Eq.~\eqref{eq:expansion1} can now be solved to any order $j$, the solutions become increasingly cumbersome. Up to here, we have assumed that the amplitude of the corrugations is much smaller than the length-scale on which they vary (i.e., $\partial_z R(z) \ll 1$). In order to gain analytical insight into the contributions of the orders of $Q$ larger than zero, we further assume that the amplitude is small compared with the mean value $|R(z)-R_0| \ll R_0$, so that 
\begin{equation}
\frac{|\alpha_i|}{\alpha_0} \ll 1 \text{   , for all } i > 0.
\label{eq:shallowCorrugation}
\end{equation}
Accordingly, in Eq. ~\eqref{eq:Q_expanded} we retain only those terms which are proportional to $\alpha_i$. To this end, we must calculate the contribution of zeroth order in $\alpha_i$ of the quantities $\xi_{B,i}$. Indeed
\begin{equation}
    \xi_B(z) = \pi R(z)^2 \hat{\xi}_B(z) = \pi \left[\alpha_0 + \SUMIGRZERO \alpha_i \cos(k_i z)\right]^{-1} \hat{\xi}_B(z) \approx \frac{\pi}{\alpha_0} \hat{\xi}_B(z) + \mathcal{O}(\alpha_{i>0}),
\end{equation}
and it is useful to obtain the Fourier coefficients
\begin{equation}
   \hat{\xi}_B(z) = \hat{\xi}_{B,0} + \SUMIGRZERO \hat{\xi}_{B,i} \cos(k_i z), i>0.
\end{equation}
Equation~\eqref{eq:Q_expanded} now returns the series
\begin{equation}
Q = - \frac{\curlyL}{2\beta \eta} \sum\limits_{i>0} \frac{\alpha_i}{\alpha_0} \frac{k_i}{\chi + Dk_i^2} \frac{\pi}{\alpha_0} \hat{\xi}_{B,i} \sum\limits_{j>0} (-1)^{j-1}\left( \frac{Q}{Q^*_i} \right)^{2j-1}, 
 \label{eq:dimensionlessQ}
\end{equation}
where
\begin{equation}
Q^*_i =  \frac{\pi(\chi + Dk_i^2)}{\alpha_0 k_i}.
\end{equation}
Equation~\eqref{eq:dimensionlessQ} is an implicit equation for $Q$ which always has the solution $Q=0$. This is the non-pumping, fore-aft symmetric steady state. This solution may be unstable, but it always exists due to the intrinsic fore-aft symmetry of the problem. Solutions with $Q\neq 0 $ satisfy the expression
\begin{equation}
1 = - \frac{\curlyL}{2\beta \eta} \sum\limits_{i>0} \frac{\hat{\xi}_{B,i} \alpha_i}{\alpha_0} \left( \frac{k_i}{\chi + Dk_i^2} \right)^2 \sum\limits_{j>0} (-1)^{j-1}\left( \frac{Q}{Q^*_i} \right)^{2(j-1)}.
 \label{eq:dimensionlessQ1}
\end{equation}
The sum over $j$ on the right-hand side of Eq.~\eqref{eq:dimensionlessQ1} is an alternating series which converges if and only if
\begin{equation}
\label{eq:condQ}
 \frac{|Q|}{Q^*_i} < 1.
\end{equation}
If Eq.~\eqref{eq:condQ} does not hold for all $i \in \mathbb{N}$, Eq.~\eqref{eq:dimensionlessQ1} has no solution. In that case, only the non-pumping state $Q=0$ is the solution of Eq.~\eqref{eq:dimensionlessQ}. If Eq.~\eqref{eq:condQ} is valid for all $i \in \mathbb{N}$, one obtains
\begin{equation}
 1 = - \frac{\curlyL}{2\beta \eta} \sum\limits_{i>0}  \frac{\hat{\xi}_{B,i} \alpha_i}{\alpha_0}  \left( \frac{k_i}{\chi + Dk_i^2} \right)^2 \left[ 1 + \left(\frac{Q}{Q_i^*} \right)^2 \right]^{-1}.
 \label{eq:dimensionlessQ2}
\end{equation}
In the current study, we focus on the transition between pumping and non-pumping steady states, the condition of which is given by the limit $Q \rightarrow 0$ of Eq.~\eqref{eq:dimensionlessQ2} (which always satisfies the condition given by Eq.~\eqref{eq:condQ}):
\begin{equation}
 1 = -\frac{\curlyL}{2\beta \eta} \sum\limits_{i>0}  \frac{\hat{\xi}_{B,i} \alpha_i}{\alpha_0} \left( \frac{k_i}{\chi + Dk_i^2} \right)^2 .
 \label{eq:pumpingCond}
\end{equation}
Finally, we note that the function $R(z)$ (Eq.~\eqref{eq:defR}) can be approximated up to linear order in $\alpha_i$ by a sinusoidal
\begin{equation}
R(z) \approx R_0 + \sum\limits_{i>0}R_i \cos(k_i z),
\end{equation}
where $R_0$ is the mean radius, and the coefficients $R_i$ are the amplitudes of the Fourier modes of the deviation from the average. These values are given by
\begin{equation}
R_0 = \alpha_0^{-\frac{1}{2}},
\end{equation}
and
\begin{equation}
R_i = - \frac{1}{2} \alpha_i \alpha_0^{-\frac{3}{2}}.
\end{equation}
These quantities have a more direct geometric interpretation as compared with $\alpha_0$ and $\alpha_i$, and thus allow better insight into the pumping condition (Eq.~\eqref{eq:pumpingCond}), which can be written as
\begin{equation}
0 = 1- \frac{\curlyL}{\beta \eta}  \sum\limits_{i>0} \frac{R_i}{R_0} \hat{\xi}_{B,i} \left( \frac{k_i}{\chi + Dk_i^2}\right)^2, k_i=\frac{\pi}{L}i.
\label{eq:finalPumpingCond}
\end{equation}
An immediate consequence of Eq.~\eqref{eq:finalPumpingCond} is that a flat pore ($R_i = 0$) cannot pump, in agreement with previous work~\cite{Michelin2020}. We now compare the prediction for the pumping transition given by Eq.~\eqref{eq:finalPumpingCond} with the numerical results obtained from the Lattice Boltzmann simulations.

\subsection{Comparison with numerical results}

\begin{figure*}[t]
	\centering
   \includegraphics[width=0.5\textwidth]{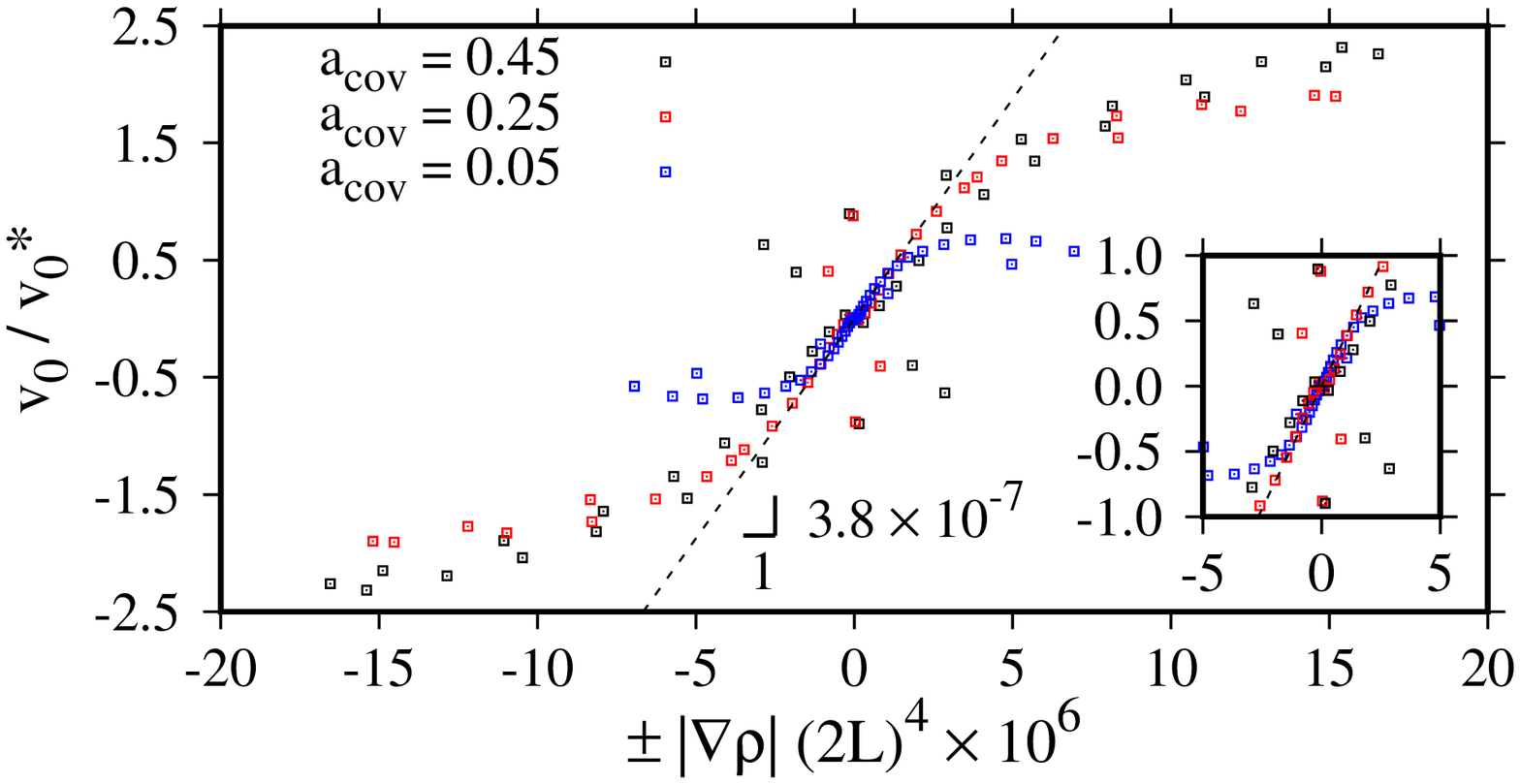}
   \caption{Normalized slip velocity $v_0 \ =|\bm{v}_0|$ as a function of the signed magnitude $\pm |\nabla \rho|$ of the gradient of the solute density along the wall. Positive (negative) values of $\pm |\nabla \rho|$ indicate that $\rho$ is locally increasing (decreasing) upon a shift towards larger values of $z$. The normalization is taken with $v_0^* = (2L)/\tau_f$, where $\tau_f=(2L)^2/\nu$ is the relaxation time of the fluid. The slope of the dashed black line equals $-\Big(\mathcal{L}/(\beta \eta)\Big)\tau_f/(2L)^5$. In lattice units, we have chosen $L=20$, $\eta = 1/6$, $U_0 = 4 \times 10^ {-4}$, $l=4$, $\xi=1$, $\chi=10^{-3}$, $\beta=1$, and $\theta = \pi/6$. The simulation box comprises $80\times80\times40$ lattice units and $Pe=1.6$.}
    	\label{mobility}
\end{figure*}
In order to judge the performance of the approximate, analytical theory derived above, we confront it with the corresponding numerical results. To this end, one has to infer from the numerical simulations the values of the parameters, which enter into the analytical model. Accordingly, one first has to extract the phoretic mobility $\mathcal{L}/(\beta \eta)$ from the simulation data. For that purpose, one runs simulations which lead to steady states with $Q=0$. From these steady states, the profiles of $\rho$ and of the component of the flow velocity parallel to the wall are extracted. Averaging over a narrow spatial region, in which the potential $W$ is non-zero, renders an effective solute density at the pore wall, as well as an effective slip velocity (both as a function of $z$). From these averages, one obtains $|\bm{v}_0|$ and $|\nabla_{||}\rho|$, which are shown in Fig.~\ref{mobility}. A linear relationship emerges for small values of $|\nabla_{||}\rho|$, from which $\mathcal{L}/(\beta \eta)$ follows by using Eq.~\eqref{eq:vslip}. The proportionality constant is thus measured to be $\big( \mathcal{L}/(\beta \eta) \big) \times [\tau_f/(2L)^5] = -3.8 \times 10^{-7}$ ($=-4 \times 10^{-3}$ in lattice units), and is used for the comparison between the simulation data and the calculations. This linear relationship follows from the Anderson approach~\cite{Anderson1989}, which assumes local thermal equilibrium along the radial direction, zero flux of solute through the wall, and a flat pore. Our simulations go beyond these assumptions, which may explain the appearance of non-linear deviations in the simulation data. The need to perform a surface-average over a non-axially symmetric steady state may induce further scattering of the data.
\\
Since the simulations make use of a source term located at the wall, and the calculations employ a source in the bulk, bridging the gap between these two source terms is not obvious. Since the flow is completely fixed upon the slip velocity, which in turn is entirely determined by $\rho$ near the wall, we pick the source term for the calculations such that the production of solute near the wall is the same in both the calculations and the simulations. In the simulations, the number of solute molecules synthesized in a lattice unit adjacent to the catalytic section of the wall is $\xi_{sims} (\Delta x)^2$, where $\xi_{sims}$ is the source constant in Eq.~\eqref{eq:source_sims} and $\Delta x$ is the length of the lattice unit. Within the analytic approach, the number of solute molecules produced in a lattice unit by the bulk source is $\hat{\xi}_B (\Delta x)^3$ in Eq. \eqref{eq:adv-diff}. By using Eq.~\eqref{eq:source},  we obtain the connection between the functional form of the source terms in the simulations and in the analytical approach, respectively:
\begin{equation}
   \hat{\xi}_B(z, a_{cov})=\xi_{sims}(z, a_{cov}) (\Delta x)^{-1}\,,
\end{equation}
where in the simulations $a_{cov}$ is the fraction of the pore covered by catalyst. That expression for $\hat{\xi}_B$, which provides a bridge between the simulations and the calculations, results in an infinite sum on the right-hand-side of Eq.~\eqref{eq:finalPumpingCond}. This equation can be solved numerically for the critical value of the diffusion coefficient $D_c$ (below which there is a pumping solution) as a function of $a_{cov}$. We obtain the critical value $Pe_c$ from $D_c$ via Eq.~\eqref{eq:peclet}. For the numerical solution of Eq.~\eqref{eq:finalPumpingCond}, we include only the first fifty terms on the right-hand-side, because including further terms does not significantly alter the results. We note that the result is a polynomial equation of the order of fifty in $D^2$, and thus in principle there are fifty possible solutions. However, plotting the right-hand-side of Eq.~\eqref{eq:finalPumpingCond} for multiple values of $a_{cov}$ shows, that there is only one real solution within the parameter range probed by the simulations. Should there be more than one real solution, there will be multiple bifurcations in $Q$, rather than just one. The comparison between calculations and simulations is shown in Fig.~\ref{fig:diagram}. A remarkable agreement is found for the line of critical values of the P\'eclet number.

\section{Supplementary Video}
The video showcases the sustained oscillations observed in the velocity profiles for $Pe=8.0$ and $a_{cov}=0.4$. The upper panel shows the plane $x=0$, while the lower panel shows the plane $z=0.05(2L)$. The parameters are $R_{max}/(2L)=1$, $\nu \tau_f/(2L)^2= 1$, $\beta U_0 = 4 \times 10^ {-4}$, $l/(2L)=0.1$, $\xi (2L)^2\tau_f=1.5 \times 10^7$, and $\chi \tau_f = 9.6$. In lattice units: $L=20$, $\eta = 1/6$, $U_0 = 4 \times 10^ {-4}$, $l=4$, $\xi=1$, $\chi=10^{-3}$, $\beta=1$, and $\theta = \pi/6$. The simulation box is of size $80\times80\times40$.

\end{document}